# Microcomb spectral fingerprinting enables parts-per-trillion methane detection with low humidity cross-sensitivity

Shiyan Cao[1], Yinchao Xu[1], and Judith Su[1,2]

[1]Wyant College of Optical Sciences, The University of Arizona, Tucson, AZ, USA

[2] Department of Biomedical Engineering, The University of Arizona, Tucson, AZ, USA

*judy@optics.arizona.edu

Optical frequency combs have transformed molecular spectroscopy by enabling broadband, high-resolution analysis, yet they are conventionally used as light sources to interrogate external samples. Here, we introduce an integrated microcomb sensing architecture that unifies resonantly enhanced evanescent sensing with multiwavelength molecular fingerprinting in a single ultrahigh-quality-factor silica microresonator. Gaseous analytes interact directly with the evanescent field of light recirculating within the resonator, enhancing sensitivity while encoding analyte-induced perturbations across the microcomb's collective spectral response. Rather than relying on a single optical resonance, we exploit distributed responses across multiple comb lines to generate multidimensional molecular fingerprints. Adding more comb lines systematically improves analyte discrimination, showing that the microcomb provides sensing information beyond what individual spectral channels offer. Using methane as a model analyte, we achieve an estimated detection limit of 178 parts per trillion, with minimal spectral responses to ammonia, carbon monoxide, hexane, and water vapor at relative humidities up to 90%. By replacing physically long interaction paths with resonant light recirculation in an ultrahigh-Q microresonator, this architecture brings compact optical sensing into the parts-per-trillion regime while exploiting distributed spectral information for chemical identification.

## Introduction

Methane is a potent greenhouse gas with a global warming potential substantially greater than carbon dioxide over a 20-year timescale[1,2]. As the primary component of natural gas, methane is emitted from energy production, industrial processes, agriculture, and waste management, with significant environmental, economic, and safety implications[3]. Consequently, sensitive, selective, and field-deployable methane sensors are essential for emissions monitoring, leak detection, industrial process control, and environmental surveillance[4,5]. Detecting methane at trace concentrations, however, remains challenging because its near-infrared overtone and combination-band transitions are intrinsically weak[6,7], requiring highly sensitive measurement techniques while maintaining robustness under realistic operating conditions.

Numerous optical sensing technologies have been developed for trace methane detection. Tunable diode laser absorption spectroscopy (TDLAS)[8], cavity ring-down spectroscopy (CRDS)[9,10], cavity-enhanced absorption spectroscopy (CEAS)[11,12], photoacoustic spectroscopy (PAS)[13], and dual-comb spectroscopy[14,15] have achieved sensitivities ranging from parts-per-billion to tens of parts-per-trillion by directly interrogating molecular absorption spectra. These approaches have established powerful platforms for quantitative molecular spectroscopy, but their high sensitivity often relies on long optical interaction lengths[16], stabilized laser systems[17], high-finesse cavities[18], or sophisticated optical instrumentation[19], limiting their implementation in compact integrated photonic platforms.

In parallel, compact methane sensors based on graphene, carbon nanotubes, and other functional nanomaterials[20–22] have attracted considerable interest because of their small footprint[20], low power consumption[23] and compatibility with chip-scale MEMS-based integration[24]. However, chemiresistive sensing mechanisms are intrinsically susceptible to humidity because adsorbed water molecules can compete with target gases for active surface sites, alter surface charge-transfer reactions, and change the carrier concentration and electrical conductivity of the sensing layer[25–27]. As a result, humidity often introduces baseline drift, degrades sensitivity and selectivity, or requires additional calibration and compensation strategies for reliable operation[25–28]. Optical whispering-gallery-mode (WGM) microresonators provide an attractive alternative because their sensing response arises from light–matter interactions between the resonantly confined optical field and surrounding analytes[29–31]. These interactions perturb the effective refractive index, molecular absorption, or scattering loss of the cavity and are read out optically as resonance shifts or cavity-loss changes, rather than as changes in surface electrical conductivity[29–32]. By confining light within ultrahigh-quality-factor microcavities, WGM resonators substantially enhance light–matter interactions while maintaining compact device dimensions[29–31,33]. Previous WGM sensors have demonstrated highly sensitive detection of refractive index changes, nanoparticles, biomolecules, and gases through resonance shifts or cavity-loss measurements, illustrating the potential of integrated optical resonators for chemical sensing[28–31,34].

More recently, ultrahigh-Q microresonators have emerged as efficient sources of optical frequency combs generated through Kerr and Raman nonlinear optical processes[35–40]. Although optical frequency combs have become indispensable tools for broadband molecular spectroscopy, they have most commonly been employed as broadband optical sources that interrogate external samples[41,42], with molecular information subsequently recovered through spectroscopic reconstruction[15,41,42]. Comparatively little attention has been devoted to exploiting the intrinsic spectral response of an integrated microcomb itself as the sensing observable. Rather than serving solely as a light source, a microcomb may encode molecular perturbations directly within its distributed spectral response, providing an integrated sensing modality fundamentally different from conventional frequency-comb spectroscopy.

Here we demonstrate an integrated microcomb sensing architecture based on an ultrahigh-quality-factor ($Q > 10^8$) silica microtoroid resonator that simultaneously generates a broadband optical frequency comb[43,44] and performs molecular sensing. Instead of monitoring individual optical resonances[34,45–54] or isolated absorption features, we introduce microcomb spectral fingerprinting, in which we analyze weak perturbations distributed across multiple comb lines collectively to generate an information-rich molecular fingerprint. Using methane as a model analyte, we distinguish it from nitrogen, ammonia, carbon monoxide, and hexane, and from backgrounds with relative humidities up to 90%, while achieving an estimated methane detection limit of 178 ppt. This 178-ppt detection limit brings compact microresonator-based sensing into the sensitivity regime of state-of-the-art optical methane sensors, without requiring the long free-space or fiber-based interaction paths conventionally used for parts-per-trillion detection. More broadly, this work establishes integrated microcomb sensing as a complementary approach to conventional frequency-comb spectroscopy and provides a general framework for compact, information-rich molecular sensing using nonlinear photonic microresonators.

## Results and Discussion

### *Integrated microcomb sensing platform for molecular detection*

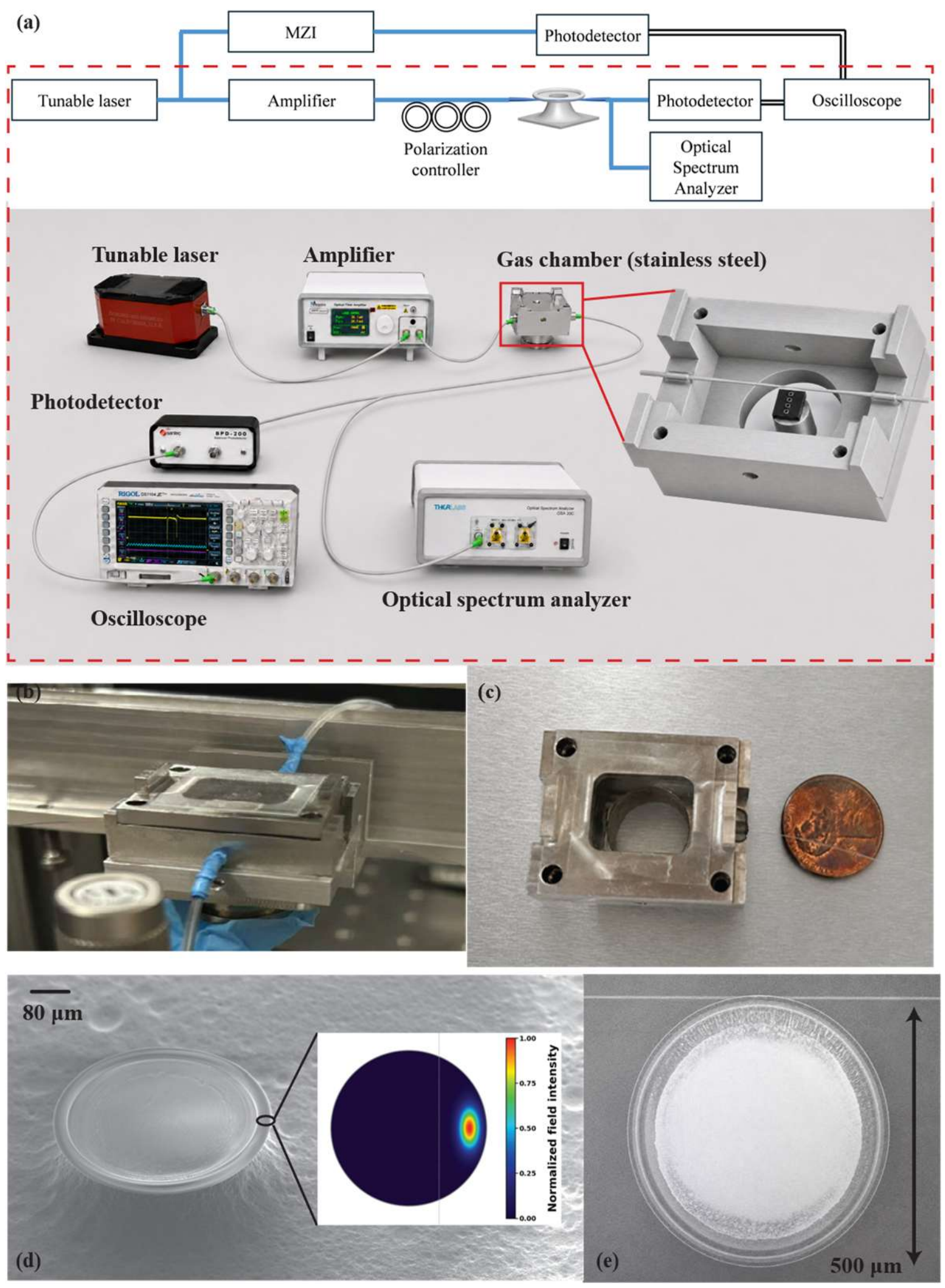

**Fig. 1. Microtoroid frequency-comb platform for methane sensing.** (a) Experimental setup. A continuous-wave laser (1064–1066 nm) is amplified to 600 mW and coupled into a silica microtoroid via a tapered optical fiber. The fiber–microtoroid coupling region is enclosed in a custom gas chamber, with an oscilloscope and optical spectrum analyzer monitoring the output. Inset: Chamber design showing fiber supports, gas inlet/outlet, and nanopositioner-

mounted microtoroid for precise coupling alignment. (b) Photograph of the sealed gas chamber with inlet and outlet tubing. (c) Gas chamber shown with a U.S. quarter for scale. (d) SEM image of a 500-μm-diameter microtoroid and simulated fundamental optical mode. (e) Microscope image of fiber–microtoroid coupling.

To investigate whether an optical microcomb can function as a distributed molecular sensor, we developed an integrated sensing platform based on an ultrahigh-quality-factor ($Q > 10^8$) silica microtoroid resonator that simultaneously generates a broadband optical frequency comb and interacts with gaseous analytes through its evanescent field (Fig. 1a). Resonant recirculation within the ultrahigh-Q cavity enables repeated interaction between the optical field and molecules near the resonator surface, enhancing analyte-induced perturbations. Rather than monitoring a single resonance or wavelength, the platform captures these perturbations across the microcomb's collective spectral response, providing a multidimensional sensing observable.

A continuous-wave laser operating near 1065 nm was coupled into the microtoroid through a tapered optical fiber, generating a Raman-assisted optical frequency comb spanning more than 200 nm through cavity-enhanced nonlinear optical interactions (Fig. 1a). The resonator exhibited an optical quality factor of approximately $1.5 \times 10^8$ (Supplementary Fig. 1), providing long intracavity photon lifetimes and efficient nonlinear frequency conversion. The resulting microcomb comprises numerous discrete spectral lines whose distributed responses provide substantially richer information than a single-wavelength measurement.

For controlled gas measurements, we enclosed the fiber–microtoroid coupling region in a sealed gas chamber while maintaining stable optical coupling (Fig. 1b–c). We permanently fixed a tapered fiber with ~90% transmission efficiency within the chamber and mounted the microtoroid on a precision translation stage to enable reproducible alignment. This configuration maintained stable evanescent coupling while allowing controlled delivery of methane, water vapor, and chemical interferents to the sensing region.

The operating principle differs from conventional frequency-comb spectroscopy, in which a frequency comb serves primarily as a broadband light source for interrogating an external sample. Here, gaseous analytes interact directly with the evanescent field of resonantly recirculating light and perturb the microcomb generated within the sensing resonator itself. The resulting changes are distributed across multiple comb lines, producing an analyte-dependent spectral fingerprint that reflects the nonlinear optical system's collective response.

*Simulated distributed spectral responses establish the sensing principle*

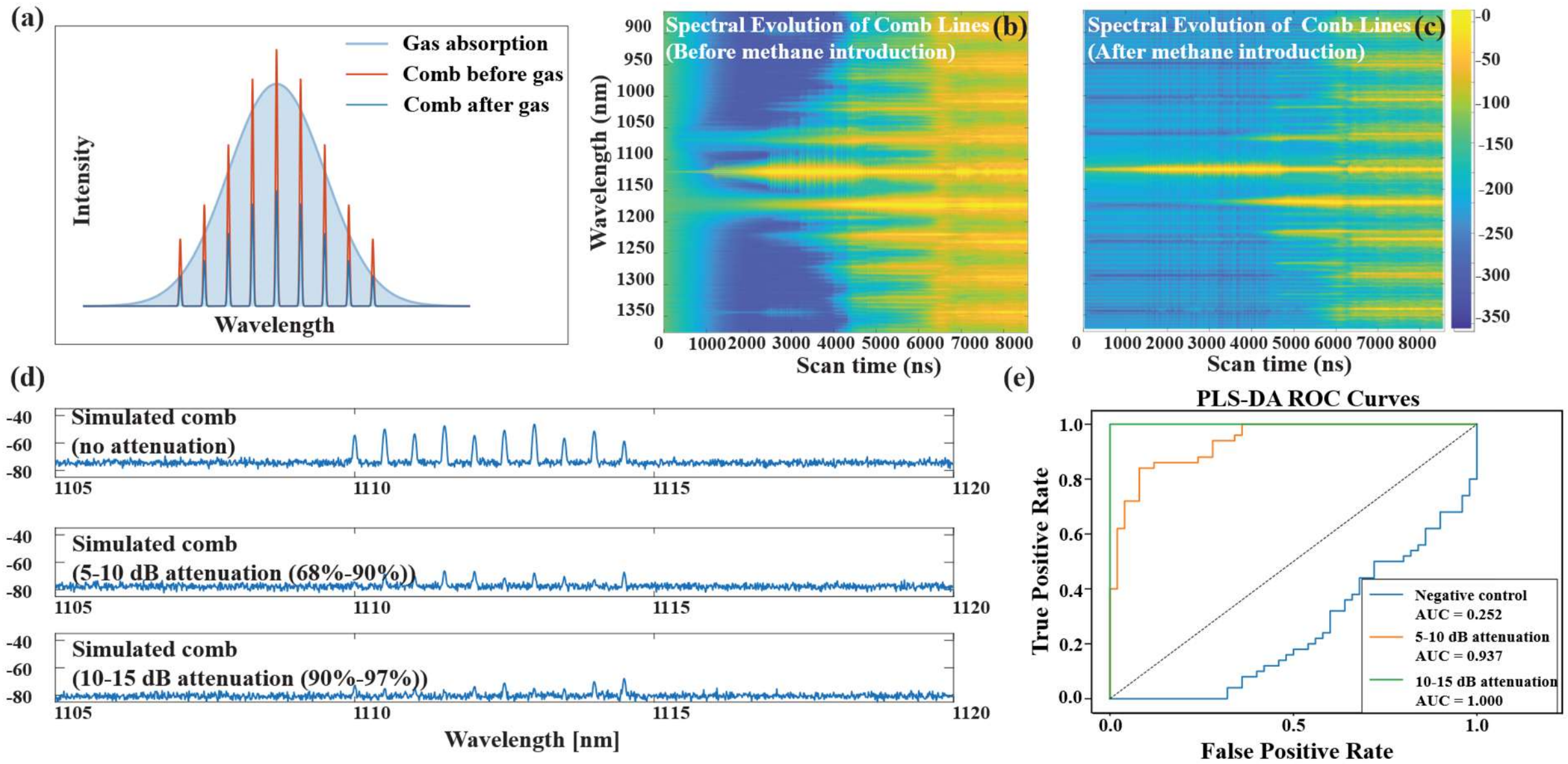

**Fig. 2. Numerical simulation of methane-induced frequency-comb attenuation and classification using partial least-squares discriminant analysis (PLS-DA).** (a) Schematic illustration of intensity-based gas sensing through wavelength-dependent attenuation of optical frequency-comb lines. Methane absorption features overlap with the generated comb spectrum in the 1100–1200 nm region. (b) Simulated spectral evolution of the frequency comb in the absence of methane, obtained by solving the Lugiato–Lefever equation (LLE). The normalized pump–cavity detuning is linearly swept from $\Delta_i = -2$ to $\Delta f = 14$ over 8580 ns (120 photon lifetimes). (c) Corresponding spectral evolution in the presence of methane, modeled by introducing an additional wavelength-dependent intracavity loss term. (d) Simulated comb-line spectra in the 1110–1115 nm region under three attenuation conditions: no attenuation, 5–10 dB attenuation (68%–90% intensity reduction), and 10–15 dB attenuation (90%–97% intensity reduction). Randomized attenuation is applied to emulate concentration-dependent gas absorption. (e) Receiver operating characteristic (ROC) curves from partial least-squares discriminant analysis (PLS-DA) classification of the simulated comb-line spectra. The area under the curve (AUC) values are 0.257, 0.937, and 1.000 for the unattenuated comb, low-attenuation condition, and high-attenuation condition, respectively, with corresponding classification accuracies of 50%, 88%, and 100%.

Before experimentally evaluating methane detection, we investigated whether weak molecular perturbations distributed across a broadband optical microcomb could produce a measurable and discriminable spectral signature. Rather than monitoring individual wavelengths independently, we hypothesized that analyte-induced perturbations across multiple comb lines could collectively encode a molecular fingerprint.

Methane shows multiple absorption features across the spectral range spanned by the Raman-assisted microcomb (Fig. 2a), providing wavelength-dependent molecular interaction across the generated comb. To examine how gas absorption perturbs the microcomb response, we modeled intracavity comb generation using a Raman-modified Lugiato–Lefever equation (LLE), in which the intracavity field evolves under cavity loss, pump-resonance detuning, chromatic dispersion, Kerr nonlinearity, stimulated Raman scattering, and external pumping[44,55,56]. Gas absorption was introduced through an additional loss term $G_{gas}(E)E$ , which accounts for absorption-induced attenuation within the comb spectral region. Supplementary Fig. 2 provides the full equation and numerical implementation and shows that the detuning scan increases the time-averaged intracavity power, enabling the formation of a Raman-assisted broadband comb. The simulations here predict that methane induces distributed intensity perturbations across multiple comb lines while largely preserving the overall comb envelope (Fig. 2b, c). Thus, rather than producing a single dominant sensing feature, the molecular interaction generates a characteristic pattern of spectral perturbations distributed across the microcomb.

We next tested whether this distributed response could be exploited for molecular discrimination. We generated synthetic spectra by applying concentration-dependent methane perturbations to the simulated microcomb while preserving its underlying spectral structure (Fig. 2d). We analyzed the resulting spectra using partial least-squares discriminant analysis (PLS-DA), treating the complete microcomb response as a multidimensional sensing observable.

Classification performance increased systematically with the magnitude of the methane-induced perturbation (Fig. 2e). Unperturbed spectra yielded performance consistent with random classification, whereas progressively stronger perturbations produced increasingly separable classes and improved receiver operating characteristic performance. Importantly, the simulations show that individually weak perturbations distributed across the microcomb can be integrated computationally into a robust molecular signature.

These results establish the central principle underlying microcomb spectral fingerprinting: molecular interactions need not produce a single dominant spectral feature to be detected. Instead, weak perturbations distributed across multiple comb lines collectively encode an analyte-dependent fingerprint that multivariate analysis can extract. We next test this prediction experimentally using methane as a model analyte.

## Methane produces reproducible perturbations across the optical microcomb

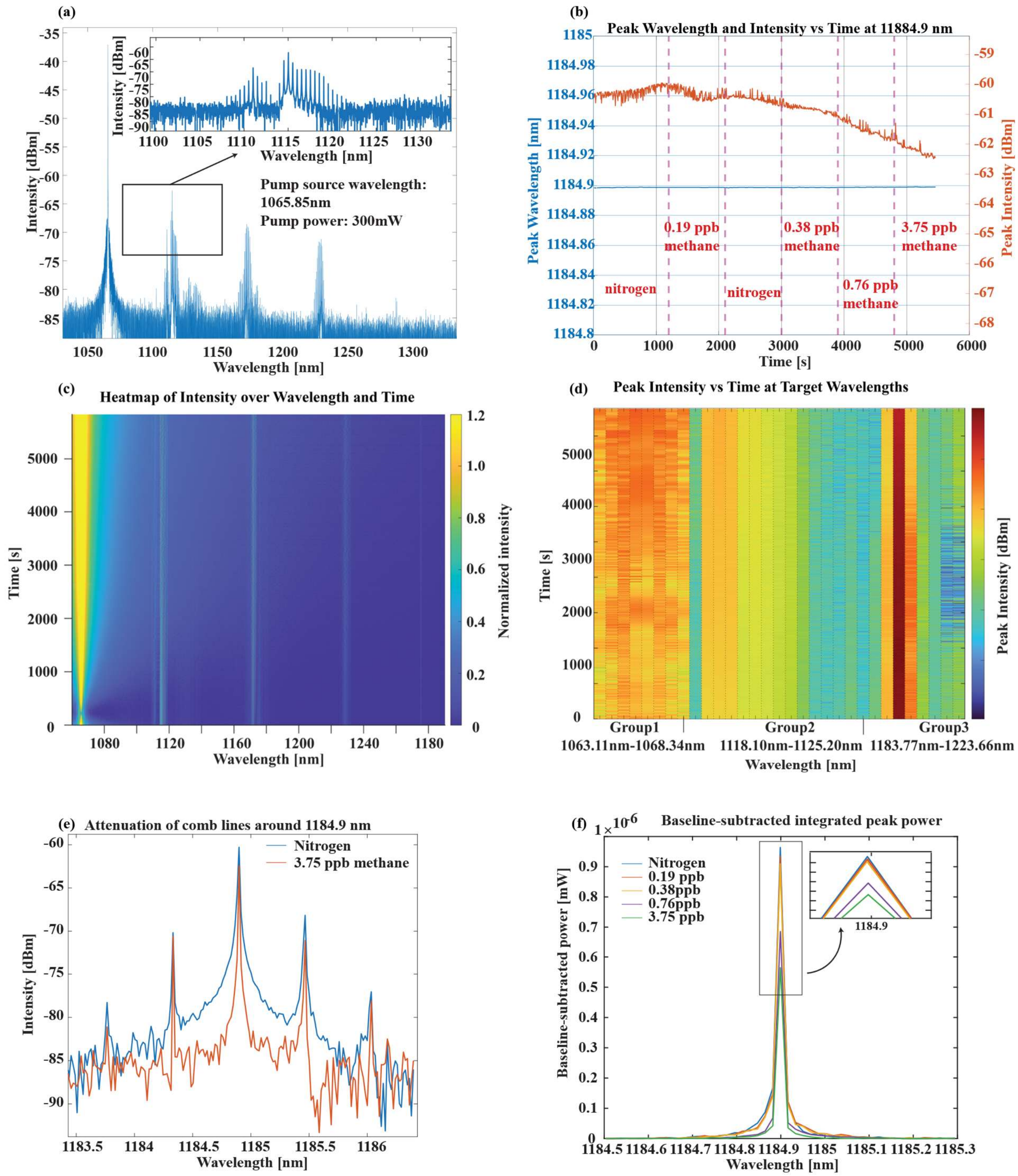

**Fig. 3 Microcomb generation and spectral response to methane.** (a) Raman frequency comb spanning 1100–1300 nm. (b) Temporal evolution of the wavelength and intensity of a representative comb line at 1184.9 nm during sequential exposure to $N_2$, 0.19 ppb $CH_4$, $N_2$, 0.38 ppb $CH_4$, 0.76 ppb $CH_4$, and 3.75 ppb $CH_4$. (c) Temporal evolution of the full microcomb spectrum over a 100 min monitoring period, showing how the intensities of multiple comb lines change simultaneously across the measured wavelength range. (d) Time-resolved heatmap of comb-line intensities

across three spectral regions, illustrating the temporal evolution of the microcomb's distributed spectral response during methane exposure (e) Comparison of five representative comb lines centered around 1184.9 nm and measured under $N_2$ and 3.75 ppb $CH_4$, showing a systematic reduction in comb-line intensity upon methane exposure. (f) Baseline-subtracted linear-power spectra of the representative comb line near 1184.9 nm for $N_2$ and increasing $CH_4$ concentrations (0.19, 0.38, 0.76, and 3.75 ppb). Subtracting the local spectral baseline suppresses the broadband pedestal and allows more direct comparison of methane-induced comb-line attenuation across concentrations.

To experimentally evaluate the integrated microcomb sensing architecture, we generated broadband Raman-assisted optical frequency combs in an ultrahigh-Q silica microtoroid and continuously monitored their spectral response during controlled methane exposure. The resulting microcomb extended from approximately 1100 to 1230 nm and comprised numerous discrete spectral channels available for simultaneous molecular sensing (Fig. 3a).

Methane exposure produced reproducible, wavelength-dependent intensity changes across multiple comb lines, while their spectral positions remained essentially unchanged (Fig. 3b). Under the conditions investigated here, the dominant sensing signature was therefore a redistribution of optical power across the microcomb rather than measurable shifts in resonance frequency. Individual comb lines exhibited relatively modest responses, but the magnitude and direction of these changes varied systematically across the spectrum, collectively generating a characteristic methane-dependent spectral pattern.

Time-resolved measurements of representative comb lines revealed a reproducible response to methane exposure (Fig. 3c). Following methane introduction, multiple comb-line intensities evolved consistently and remained stable during exposure. After methane removal, the spectral response recovered toward its initial state, demonstrating reversibility under the experimental conditions. Similar behavior across the broader microcomb spectrum (Fig. 3d) confirms that methane perturbs the microcomb's collective spectral state rather than producing an isolated response at a single wavelength.

To further examine the spectral origin of this response, we compared a group of neighboring comb lines around 1184.9 nm under nitrogen and 3.75 ppb methane exposure (Fig. 3e). Methane produced a consistent attenuation across multiple adjacent comb lines, confirming that the sensing response extends over a finite spectral region rather than being confined to a single wavelength. The raw spectra also exhibited an apparent narrowing of the comb lines following methane exposure, which is not expected from the underlying absorption response and therefore suggested an influence from the wavelength-dependent spectral background. To isolate the actual change in comb-line power, we applied a local baseline correction, removing the background contribution and eliminating the apparent linewidth variation. The resulting baseline-corrected spectra enable a more direct comparison of the concentration-dependent attenuation near 1184.9 nm (Fig. 3f). Details of the baseline correction are provided in Supplementary Fig. 3. Control experiments further established that this distributed signature was associated with methane exposure rather than measurement drift or analysis-induced separation. Identical comparisons of nitrogen measurements against independent nitrogen measurements yielded classification performance consistent with random assignment, whereas methane exposure generated reproducible spectral differences. Thus, the multivariate analysis does not intrinsically separate nominally equivalent baseline spectra.

Together, these measurements experimentally validate the distributed sensing behavior predicted by the numerical model. Methane exposure modifies the microcomb's intensity distribution while largely preserving its overall spectral envelope and comb-line positions. The sensing information is therefore encoded not in a single dominant spectral feature, but in the collective pattern of small, wavelength-dependent perturbations distributed across multiple comb lines.

This distributed response underpins microcomb spectral fingerprinting. By integrating information across multiple spectral channels, the architecture transforms individually modest optical perturbations into an information-rich molecular signature. We next examine how this collective response enhances molecular discrimination and enables parts-per-trillion methane detection.

## Collective analysis of the optical microcomb enables trace methane detection

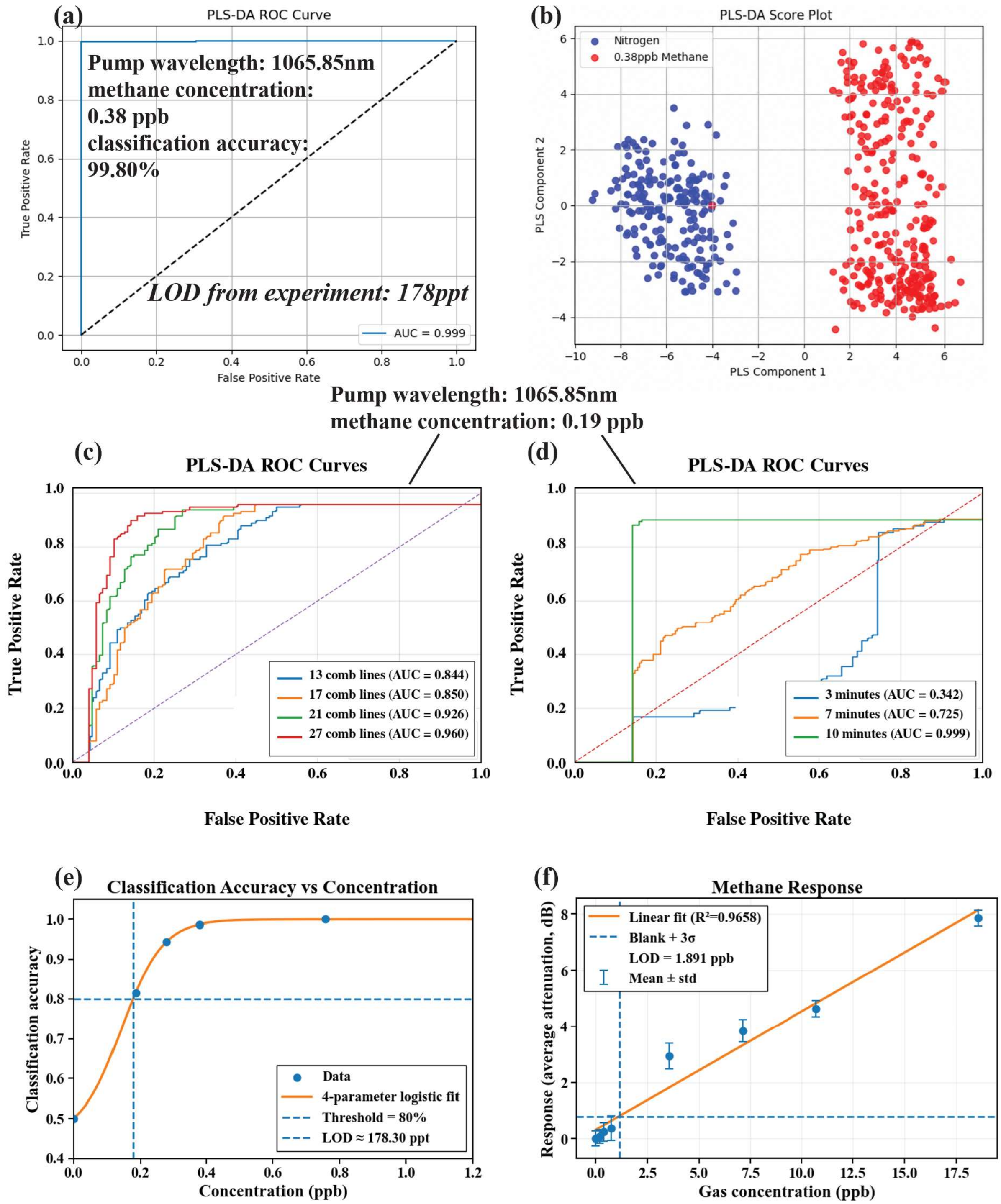

**Fig. 4 PLS-DA analysis and LOD** (a) Receiver operating characteristic (ROC) curve obtained from partial least-squares discriminant analysis (PLS-DA) of spectra acquired under nitrogen and 0.38 ppb methane exposure, with 10 min of data collected for each condition. (b) Corresponding PLS-DA score plot showing the separation between nitrogen and 0.38 ppb methane spectra in the latent-variable space. (c) PLS-DA ROC curves obtained by using different numbers of comb lines for methane classification at a pump wavelength of 1065.85 nm and a methane concentration of 0.1938 ppb. (d) PLS-DA ROC curves obtained at different spectral acquisition times, illustrating the influence of measurement duration on methane classification performance. (e) Classification accuracy as a function

of methane concentration (0.19, 0.28, 0.38, 0.76 ppb), fitted using a four-parameter logistic function. An 80% classification-accuracy threshold was used to determine the PLS-DA-based limit of detection of 178 ppt. (f) Methane response as a function of concentration from approximately 0.20 to 18.60 ppb. The response was defined as the baseline-corrected average comb-line attenuation, and error bars represent the standard deviation. The response increases linearly with methane concentration, and the intersection between the linear fit and the blank + 3σ threshold yields a signal-response-based LOD of approximately 1.89 ppb.

Having established that methane produces reproducible perturbations across multiple comb lines, we next tested whether integrating these distributed responses enhances molecular discrimination. Rather than evaluating individual wavelengths independently, we analyzed the complete microcomb spectrum using partial least-squares discriminant analysis (PLS-DA)[57,58], which projects multichannel spectral data onto latent variables that maximize separation between experimental conditions. As an additional benchmark, we compared PLS-DA with unsupervised principal component analysis (PCA). PLS-DA produced substantially clearer separation between the nitrogen and methane spectral fingerprints, supporting the use of supervised multivariate discrimination for the subsequent sensing analysis (Supplementary Fig. 4).

Using 10 min of spectral data acquired under nitrogen and 10 min under 0.38 ppb methane exposure, the PLS-DA model achieved an area under the curve (AUC) of 0.999 and a classification accuracy of 99.80%, demonstrating robust discrimination between the two conditions (Fig. 4a). The corresponding PLS-DA score plot further illustrated this discrimination, showing clear separation between nitrogen and methane spectra in the latent-variable space (Fig. 4b). The AUC evaluates the ability of the model to distinguish methane-exposed spectra from nitrogen spectra over all classification thresholds, while the classification accuracy represents the fraction of spectra assigned to the correct class. Therefore, the near-unity AUC and nearly 100% classification accuracy confirm that the distributed microcomb response contains sufficient information to identify methane at sub-ppb concentrations. In contrast, independent nitrogen-versus-nitrogen control measurements yielded classification performance consistent with random assignment (Supplementary Fig. 5), confirming that the observed separation originates from methane-induced spectral perturbations rather than measurement drift or intrinsic bias in the analysis.

To determine whether this performance arises from the collective microcomb response, we systematically varied the number of comb lines included in the analysis. Classification performance increased progressively as additional comb lines were incorporated (Fig. 4c), with the AUC improving from 0.844 using 13 comb lines to 0.960 using 27 comb lines. No individual wavelength dominated the response; instead, individually modest perturbations combined across multiple spectral channels to provide increasingly robust molecular discrimination. To further evaluate the contribution of each spectral channel, we calculated the variable importance in projection (VIP) scores from the trained PLS-DA model. The VIP analysis identifies which comb lines contribute most strongly to the classification, and showed that the most informative channels were distributed across multiple comb lines rather than being confined to a single wavelength (Supplementary Fig. 6). This systematic scaling with comb-line number, together with the VIP-score analysis, directly demonstrates that the sensing advantage derives from distributed spectral information encoded across the microcomb rather than from an exceptionally responsive individual wavelength.

We next examined how measurement duration influences discrimination. Classification performance increased with acquisition time as we accumulated additional spectral information while the microcomb remained stable (Fig. 4d). The near-unity AUC achieved at 10 min demonstrates highly reliable methane discrimination within a relatively short acquisition period, demonstrating trace methane detection on experimentally practical timescales.

We then quantified the methane detection threshold using two complementary analyses: multivariate PLS-DA classification and direct spectral-response calibration. Because the microcomb sensing signal is a multidimensional spectral fingerprint rather than the amplitude change of a single wavelength channel, we first evaluated detection performance from the concentration-dependent classification accuracy. For each methane concentration, PLS-DA was used to classify nitrogen- and methane-exposed spectra, and the resulting classification accuracy was used as the detection metric. We defined an accuracy threshold of 80% as the criterion for reliable detection. The concentration-dependent accuracy was fitted using a four-parameter logistic function, and the intersection between the fitted curve and the 80% threshold yielded a PLS-DA-based limit of detection of 178 ppt (Fig. 4e). To further validate the PLS-DA-based detection threshold, we narrowed the methane concentration range around the estimated LOD and repeated the classification analysis (Supplementary Fig. 7). Using the same 80% classification-accuracy criterion, methane concentrations of 0.15 and 0.17 ppb remained below the detection threshold, with classification accuracies of 69.00% and 67.25%, respectively, whereas 180 ppt methane exceeded the threshold with a classification accuracy of 88.22%. This low-concentration validation gives an experimentally observed detection threshold of approximately 180 ppt, in close agreement with the 178 ppt value obtained from the fitted classification-accuracy curve.

Although PLS-DA provides high sensitivity at low methane concentrations, the classification accuracy rapidly saturated above 90%, making it less suitable for evaluating how the sensor response evolves at higher concentrations. Therefore, we further performed methane sensing experiments over an expanded concentration range, and directly analyzed the spectral response. The response was extracted as the baseline-corrected average attenuation of selected comb lines, defined as:

$$R(C) = \frac{1}{N}\sum_{i=1}^{N}\left[\bar{I}_{i,N_2} - \bar{I}_{i,C}\right]$$

where $N$ is the number of selected comb lines, $\bar{I}_{i,N_2}$ is the average peak intensity of the $i^{th}$ comb line under nitrogen, and $\bar{I}_{i,C}$ is its average peak intensity at methane concentration, $C$.

In contrast to the saturated classification accuracy, the direct attenuation response increased approximately linearly with methane concentration over this higher concentration range (Fig. 4f).

This linear dependence confirms that the methane-induced comb attenuation increases systematically with gas concentration and demonstrates that the microcomb response can be used not only for low-concentration classification but also for quantitative concentration-dependent sensing. Using the blank response plus three times its standard deviation as the detection threshold, the intersection with the linear fit yielded a direct-response-based LOD of 1.89 ppb.

The difference between the two LOD values reflects the distinct information each analysis uses. The direct-response method evaluates the magnitude of the average spectral attenuation and therefore provides an intuitive measure of concentration-dependent absorption, particularly at higher methane concentrations where the attenuation becomes clearly observable. In contrast, PLS-DA accounts for correlations among multiple comb lines and utilizes the full multivariate spectral information. As a result, PLS-DA is more sensitive to weak but reproducible low-concentration perturbations, enabling methane detection at substantially lower concentrations than direct response analysis. The two approaches are therefore complementary: PLS-DA provides enhanced sensitivity for trace detection, while direct-response analysis more clearly captures concentration-dependent attenuation at higher methane levels.

These results establish the central advantage of microcomb spectral fingerprinting: sensitivity and discrimination arise from collective information encoded across many spectral channels rather than from a large response at any single wavelength. By integrating individually weak but reproducible perturbations across the microcomb, PLS-DA enables methane detection down to 178 ppt. In parallel, direct analysis of the baseline-corrected comb-line attenuation shows a linear concentration-dependent response at higher methane concentrations, confirming that the microcomb also provides a quantitative spectral response beyond binary classification. The systematic improvement with increasing comb-line number demonstrates that the microcomb is not merely a broadband optical source, but an information-rich sensing state whose distributed spectral response can be exploited for ultrasensitive and concentration-dependent molecular detection.

**Methane produces distinct spectral responses from common environmental interferents**

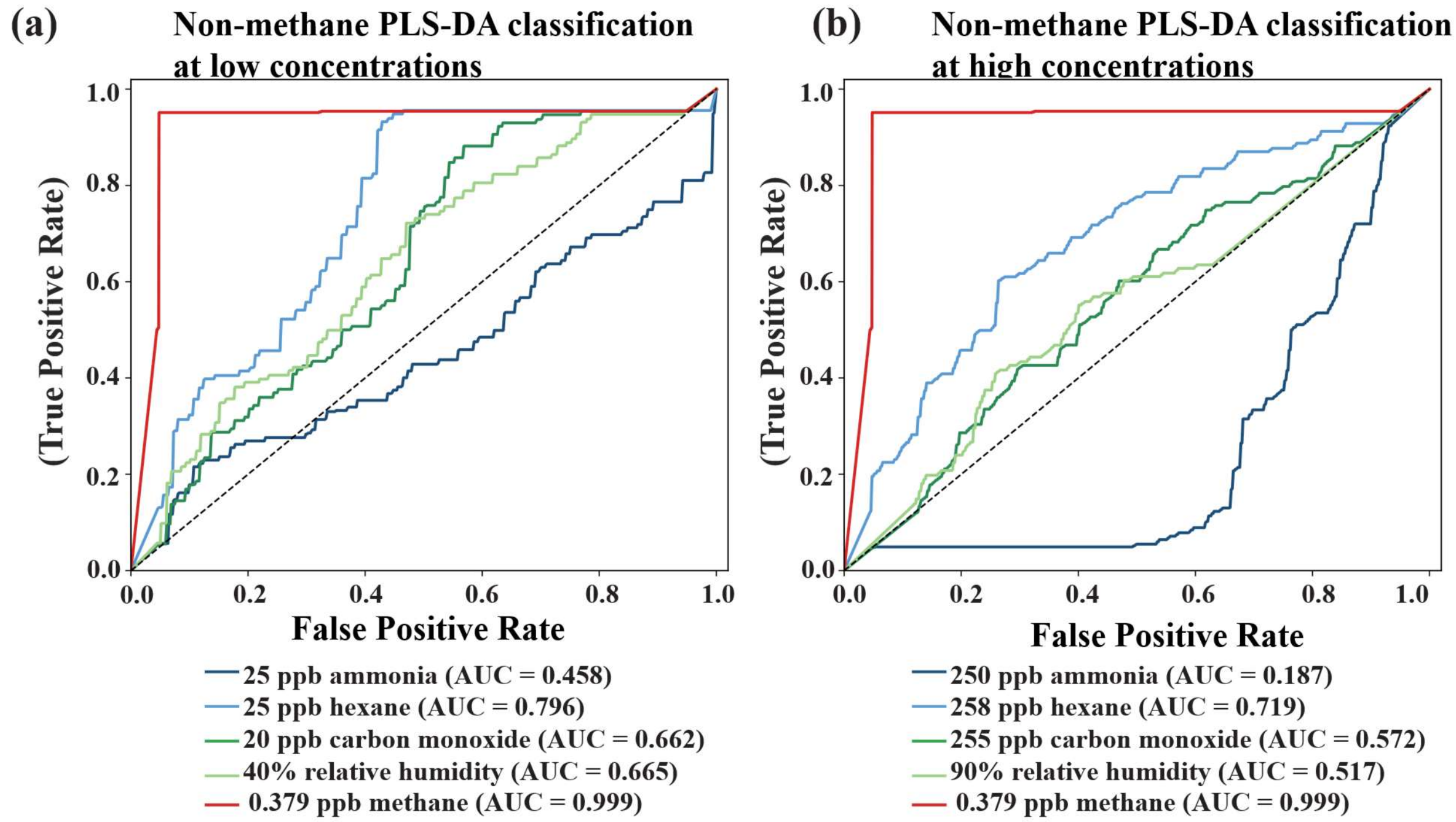

**Fig. 5. Interferent and humidity robustness of microcomb spectral fingerprinting.** (a) PLS-DA ROC curves for low-level perturbations. (b) PLS-DA ROC curves under substantially increased interferent concentrations and humidity, demonstrating robust molecular discrimination under challenging environmental conditions.

Selective molecular recognition is essential for practical methane sensing, as real-world environments contain chemical interferents and fluctuating environmental conditions that can generate nonspecific sensor responses. We therefore evaluated whether the distributed microcomb signature observed for methane could be distinguished from responses to representative chemical and environmental interferents, including ammonia, carbon monoxide, hexane, and water vapor (Fig. 5).

We analyzed the distributed spectral responses from each condition using the same multivariate framework used for methane detection. At low interferent concentrations—25 ppb ammonia, 25 ppb hexane, and 20 ppb carbon monoxide—none of the tested species produced a response that could be reliably distinguished from the nitrogen baseline (Fig. 5a). Increasing the interferent concentrations by approximately an order of magnitude, to 250 ppb ammonia, 258 ppb hexane, and 255 ppb carbon monoxide, similarly produced limited class separation (Fig. 5b). In contrast, methane at sub-ppb concentrations generated a readily distinguishable distributed spectral signature. These results demonstrate strong methane selectivity, even compared with chemically diverse interferents present at concentrations orders of magnitude higher.

Humidity presents a particularly stringent challenge for trace gas sensing because water vapor is ubiquitous and can dominate many target analytes by several orders of magnitude. We therefore examined the microcomb response at relative humidities up to 90%. Despite the large increase in water-vapor concentration, the resulting spectral response remained weakly distinguishable from the nitrogen baseline (Fig. 5b). Thus, the platform retains strong methane selectivity even under high-humidity conditions, demonstrating robustness against a major environmental interferent in practical gas sensing.

The markedly different responses to methane and the tested interferents further indicate that methane detection cannot be explained by generic changes in optical power, gas flow, or environmental perturbation. Instead, methane produces a reproducible distributed modulation across the microcomb that substantially higher concentrations of ammonia, carbon monoxide, hexane, or water vapor do not reproduce. The ability to distinguish sub-ppb methane against these competing perturbations highlights the selectivity encoded in the collective spectral response.

Together, these results demonstrate that microcomb spectral fingerprinting combines parts-per-trillion sensitivity with robust molecular selectivity. Individually modest perturbations distributed across multiple comb lines collectively enable methane detection while rejecting chemically diverse interferents and water vapor at relative humidities up to 90%. This combination of sensitivity, selectivity, and environmental robustness provides a pathway to compact molecular sensing in complex, variable environments.

**Comparison with other methods**

**Table 1. Comparison of representative approaches for trace methane detection.**

| **Technique** | **LOD (values)** | **Detection time** | **Selectivity** | **Humidity tolerance** |
|---|---|---|---|---|
| CRDS / OA-ICOS[59] | <1.5 ppb (3σ precision) | 5 s | Very high | Good, but water lines require correction |
| TDLAS / WMS[60] | 0.53 ppb | 230 s | High | Good if spectral interference is handled |
| QEPAS / PAS[61] | 1.321 ppb | 1000 s | High | Usually good, but relaxation and acoustic effects matter |
| FTIR spectroscopy[62] | 32 ppb | 40 s | High | Moderate; water background can complicate spectra |
| Dual-comb spectroscopy[63] | Sensitive: <3 ppb | 5 min | Very high | Good with spectral fitting |
| NDIR methane sensor[64] | 1 ppm | seconds-scale response | Medium–high | Good for industrial use, but water/$CO_2$ cross-effects may need compensation |
| MOS chemiresistive sensor[65] | 250 ppm | 57–210 s | Low–medium | Often problematic; temperature/humidity compensation needed |
| Electrochemical / mixed-potential sensor[66] | 9 ppm | 40 s | Medium | Humidity affects electrolyte/device behavior |
| Catalytic bead / pellistor[67] | %LEL level | <20 s | Low for methane specifically | Often robust |
| **This work** | **180 ppt** | **8 min** | **High** | **Very robust** |

## Conclusion

We have demonstrated an integrated microcomb sensing architecture that combines resonantly enhanced evanescent light–matter interactions with distributed spectral fingerprinting for ultrasensitive molecular detection. By analyzing analyte-induced perturbations collectively across multiple comb lines, the platform achieves an estimated methane detection limit of 178 ppt while exhibiting selectivity against chemical interferents and distinct responses to humidity levels up to 90%. The systematic improvement in discrimination with increasing comb-line number demonstrates the advantage of distributed spectral information over individual wavelength measurements.

This work establishes a sensing paradigm in which an optical microcomb serves not merely as a broadband light source, but as an information-rich molecular sensing element. By integrating resonant enhancement, nonlinear optical interactions, and multivariate spectral analysis within an ultrahigh-Q microresonator, this approach provides a foundation for compact, highly sensitive molecular sensors. The reproducibility of Raman-assisted microcomb generation (Supplementary Fig. 8) further supports the potential for advanced machine-learning models, extending the platform beyond binary gas detection toward quantitative concentration prediction and multi-analyte sensing. Further advances in microresonator design, comb generation, and computational analysis could extend this platform to additional analytes and complex gas mixtures, with applications in environmental monitoring, industrial process control, and chemical analysis.

## Methods

### Microtoroid fabrication

Silica microtoroid resonators were fabricated on silicon wafers with thermally grown silica using standard photolithographic patterning, buffered oxide wet etching, $XeF_2$ isotropic silicon undercut etching, and $CO_2$ laser reflow. Following undercutting, laser reflow transformed the silica microdisks into toroidal resonators with atomically smooth surfaces that support ultrahigh-quality-factor whispering-gallery modes. The resonators used in this work had outer diameters of approximately 500 µm.

## Microcomb generation and optical measurements

A Newport TLB-6700 Velocity tunable continuous-wave diode laser operating near 1065 nm was used as the pump source. The laser was coupled into the ultrahigh-Q silica microtoroid resonator through a tapered single-mode 780 nm optical fiber. Before coupling into the resonator, the laser output was amplified using an Amonics high-power ytterbium-doped fiber amplifier (AYDFA-aa-b-cc). The amplified pump light excited whispering-gallery-mode resonances in the silica microtoroid, where the enhanced intracavity field generated a broadband Raman-assisted optical frequency comb through cavity-enhanced Kerr and Raman nonlinear interactions.

The generated comb spectra were recorded using a Thorlabs optical spectrum analyzer. During wavelength tuning and resonance coupling, the cavity transmission and resonance dynamics were monitored using a Rigol DS1104 oscilloscope. The optical quality factor of each resonator was determined by fitting the calibrated cavity transmission spectrum with a Lorentzian function. We used a Mach–Zehnder interferometer as a frequency reference to calibrate the laser scan, allowing accurate extraction of the resonance linewidth and quality factor. (Supplementary Fig. 1).

## Gas delivery

The fiber–microtoroid coupling region was enclosed within a sealed gas chamber to permit controlled exposure of the evanescent optical field to gaseous analytes. Gas concentrations were generated using a FlexStream gas standard generator coupled to mass-flow controllers. Target concentrations were obtained by diluting permeation-tube source gases with nitrogen through sequential dilution stages before introduction into the sensing chamber at a constant flow rate. Nitrogen was used as the baseline gas unless otherwise specified.

## Spectral preprocessing

Optical spectra were acquired sequentially and exported as CSV files containing wavelength and intensity information. For each selected comb line, the peak intensity within a predefined wavelength window was extracted to generate a multichannel feature matrix, where rows corresponded to individual spectra and columns corresponded to comb-line intensities. Relative acquisition times were determined from file timestamps. Spectra collected during gas-transition periods were excluded from subsequent analyses to ensure that only steady-state conditions were considered.

## Multivariate analysis

### PLS-DA analysis

Partial least-squares discriminant analysis (PLS-DA) was used to analyze the gas-dependent microcomb spectral response. For each recorded optical spectrum, the intensities of selected comb lines were extracted after spectral preprocessing and used as input features. Each spectrum was assigned a class label according to the gas condition during acquisition, such as nitrogen, methane, or interfering gases. The resulting data matrix had dimensions $N \times P$, where $N$is the number of acquired spectra and $P$is the number of comb-line features included in the analysis.

For binary classification, the class labels were encoded as numerical values, with nitrogen assigned as the negative class and methane assigned as the positive class. For multi-class classification, labels were converted into a one-hot encoded response matrix. Before PLS-DA training, the feature matrix was standardized using the mean and standard deviation of the training set. The same scaling parameters were then applied to the test set to avoid information leakage. The dataset was divided into training and testing subsets using stratified sampling to preserve the class distribution. A PLS regression model was then trained using the standardized comb-line intensity matrix and the encoded class labels. The number of latent variables was selected before model evaluation and kept fixed for each comparison.

For binary PLS-DA, we used the continuous PLS prediction score as the discriminant variable. We generated receiver operating characteristic curves by sweeping the classification threshold, and calculated the area under the ROC curve to quantify class separability. We calculated classification accuracy after assigning spectra to classes using a fixed decision threshold. For multi-class PLS-DA, the predicted class was assigned according to the largest predicted class score. Score plots were generated from the first two PLS latent variables to visualize the separation of different gas conditions in the reduced-dimensionality space.

To evaluate the role of distributed spectral information, we repeated the PLS-DA analysis using different numbers of comb lines. In each case, we used the selected comb-line intensities to construct a new feature matrix and applied the same training, testing, scaling, and evaluation procedures. To evaluate the effect of acquisition time, we truncated spectra to different time windows before model training and testing, allowing classification performance to be compared as a function of measurement duration. We further performed variable importance in projection analysis to identify the comb lines contributing most strongly to gas discrimination. Comb lines with larger VIP scores were interpreted as having stronger contributions to the PLS-DA classification model.

**Numerical simulations**

Microcomb generation was modeled using a Raman-modified Lugiato–Lefever equation incorporating cavity loss, laser detuning, chromatic dispersion, Kerr nonlinearity, stimulated Raman scattering, and external pumping. Methane absorption was represented by introducing an additional wavelength-dependent loss term that attenuated comb lines overlapping the molecular absorption band. Simulated spectra before and after methane exposure were compared to evaluate the expected spectral response. Synthetic comb spectra with concentration-dependent attenuation were subsequently analyzed using the same PLS-DA framework as the experimental data to assess whether distributed spectral perturbations could support molecular discrimination.

**Acknowledgements**

This work was supported by the National Science Foundation through a CAREER award (No. 2237077) and an EAGER award (No. 1842045).

**Competing interests**

J.S. owns a financial stake in Femtorays Technologies, which develops label-free molecular sensors. The remaining authors declare no competing interests.

# Supplementary Information for Microcomb spectral fingerprinting enables parts-per-trillion methane detection with low humidity cross-sensitivity

Shiyan Cao[1], Yinchao Xu[1], and Judith Su[1,2]

[1]Wyant College of Optical Sciences, The University of Arizona, Tucson, AZ, USA

[2] Department of Biomedical Engineering, The University of Arizona, Tucson, AZ, USA

*judy@optics.arizona.edu

## Quality factor calculation

The quality factor of the silica microtoroid resonator was extracted from the linewidth of the resonance dip. In our setup, a Mach–Zehnder interferometer (MZI) was used to calibrate the frequency scale during wavelength sweeping. A triangle-wave voltage was applied to sweep the laser wavelength across the resonance. During the sweep, three signals were simultaneously recorded using a data acquisition card: the transmitted optical power from the coupled tapered fiber, the MZI signal, and the triangle-wave driving signal.

The MZI signal provided a sinusoidal calibration trace with a free spectral range of 39.934 MHz. This known frequency spacing was used to convert the time-domain resonance trace into a calibrated frequency axis. The resonance dip in the transmission spectrum was then fitted using a Lorentzian function, from which the full width at half maximum (FWHM) of the resonance was obtained. By comparing the fitted linewidth with the calibrated MZI frequency spacing, the resonance linewidth was determined in frequency units. Using this calibrated Lorentzian-fitting method, the quality factor of the microtoroid resonator was determined.

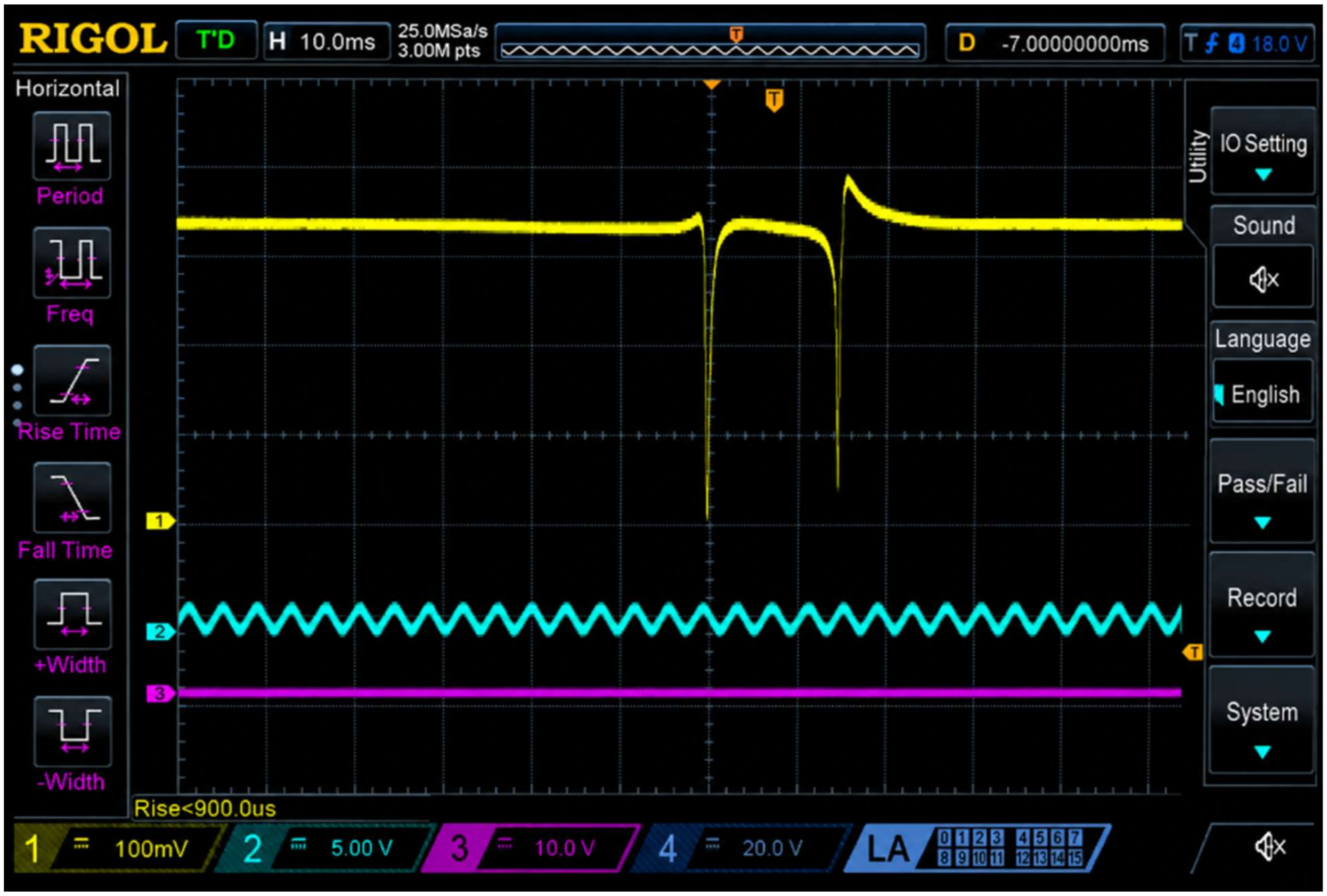

**Supplementary Figure 1.1. Signals used to measure the microtoroid resonance linewidth.** Representative oscilloscope traces showing the transmitted optical signal from the tapered-fiber–microtoroid system (yellow), the Mach–Zehnder interferometer calibration signal (cyan), and the laser-scan driving signal (magenta). The transmission dips correspond to crossings of the cavity resonance during the laser wavelength scan.

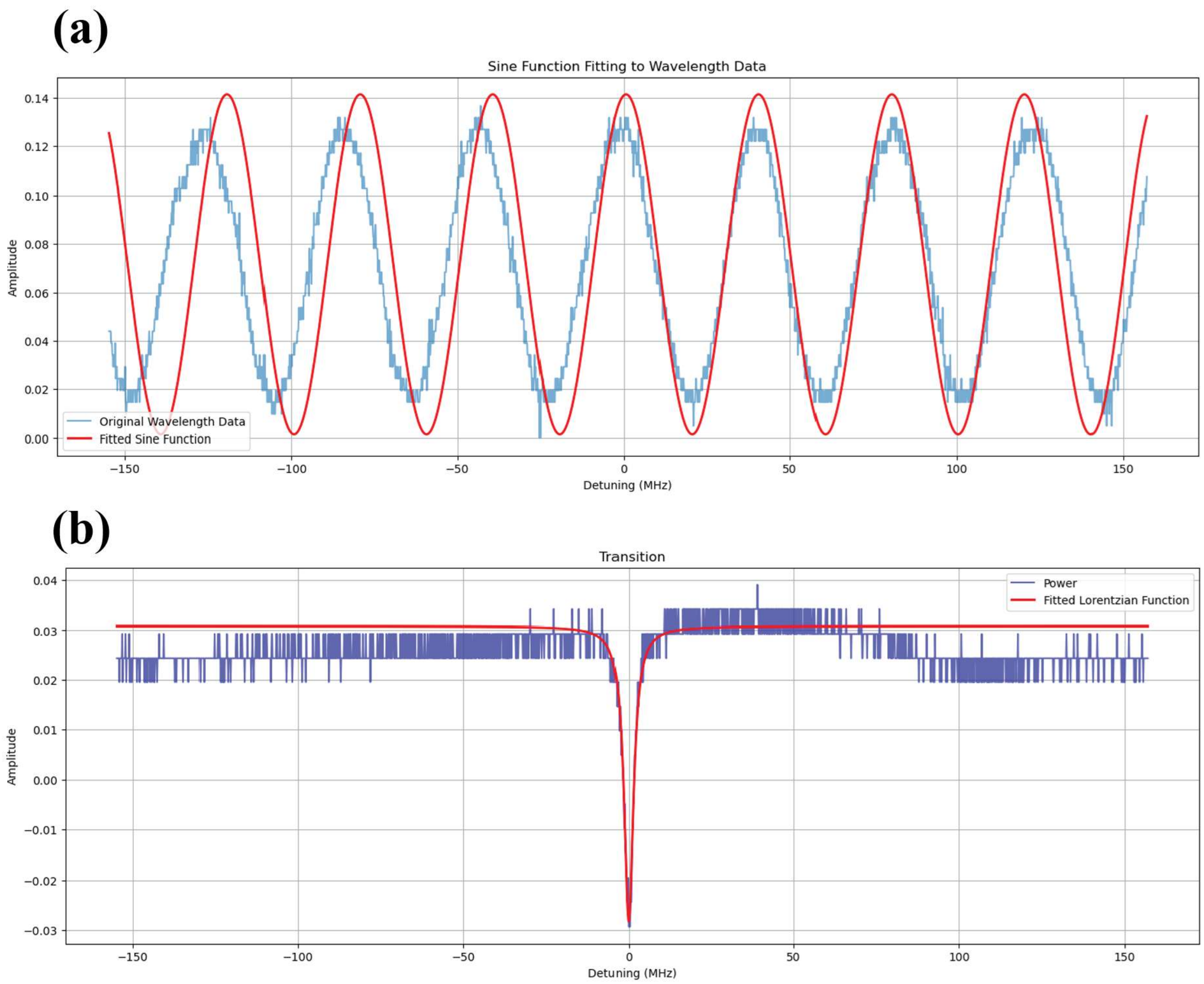

**Supplementary Figure 1.2. Frequency calibration and resonance fitting for quality-factor determination.** (a) Measured Mach–Zehnder interferometer signal (blue) and sinusoidal fit (red). The interferometer free spectral range of 39.934 MHz provides the frequency reference for calibrating the laser scan. (b) Microtoroid transmission resonance plotted against calibrated frequency detuning (blue), together with a Lorentzian fit (red). The fitted full width at half maximum, $\Delta\nu$, is used to calculate the cavity quality factor as $Q = \nu_0/\Delta\nu$, where $\nu_0$ is the resonance frequency.

pumping, and pump–resonance detuning; gas-induced absorption is excluded from the results shown.

## Raman-modified LLE

(a)

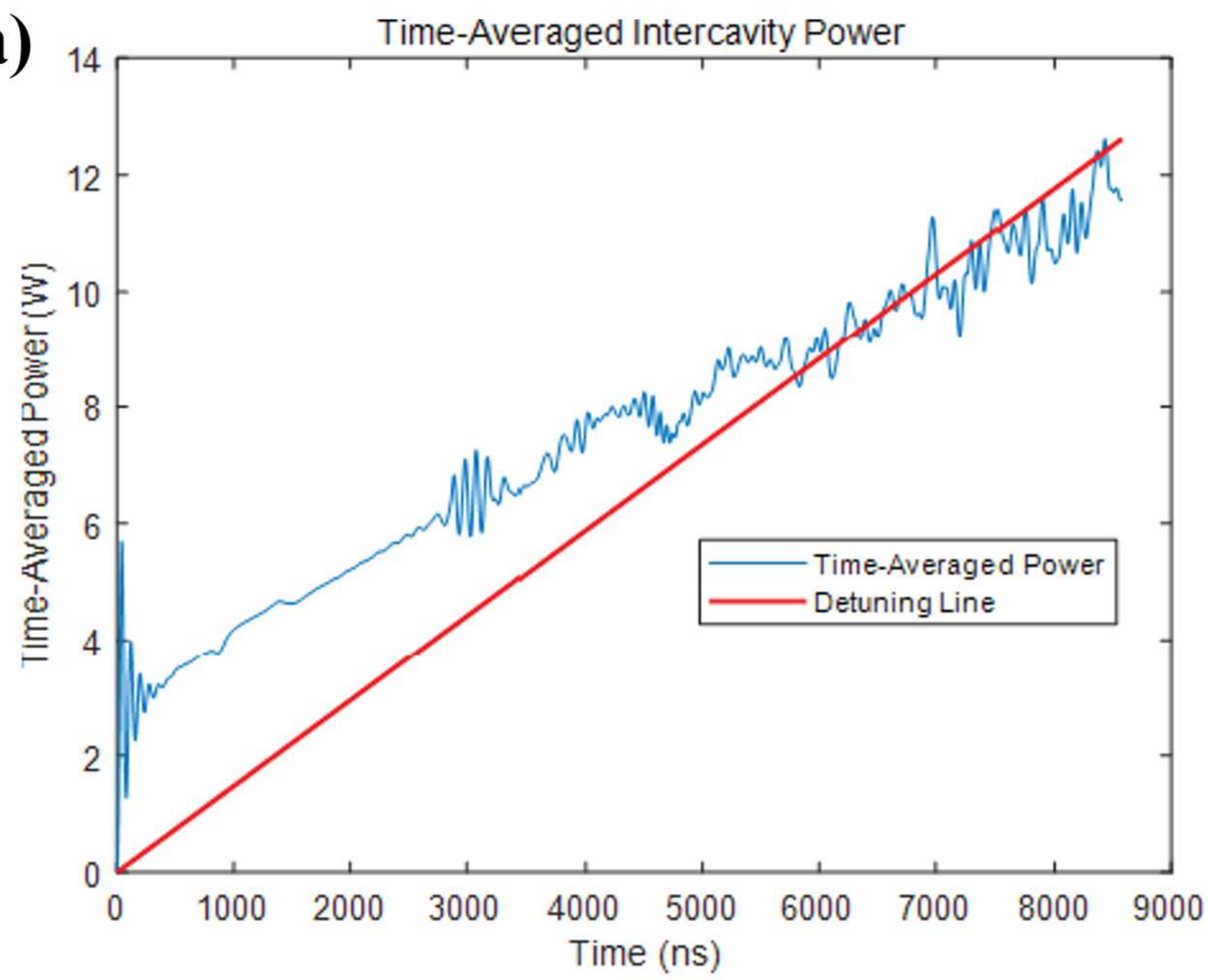

(b)

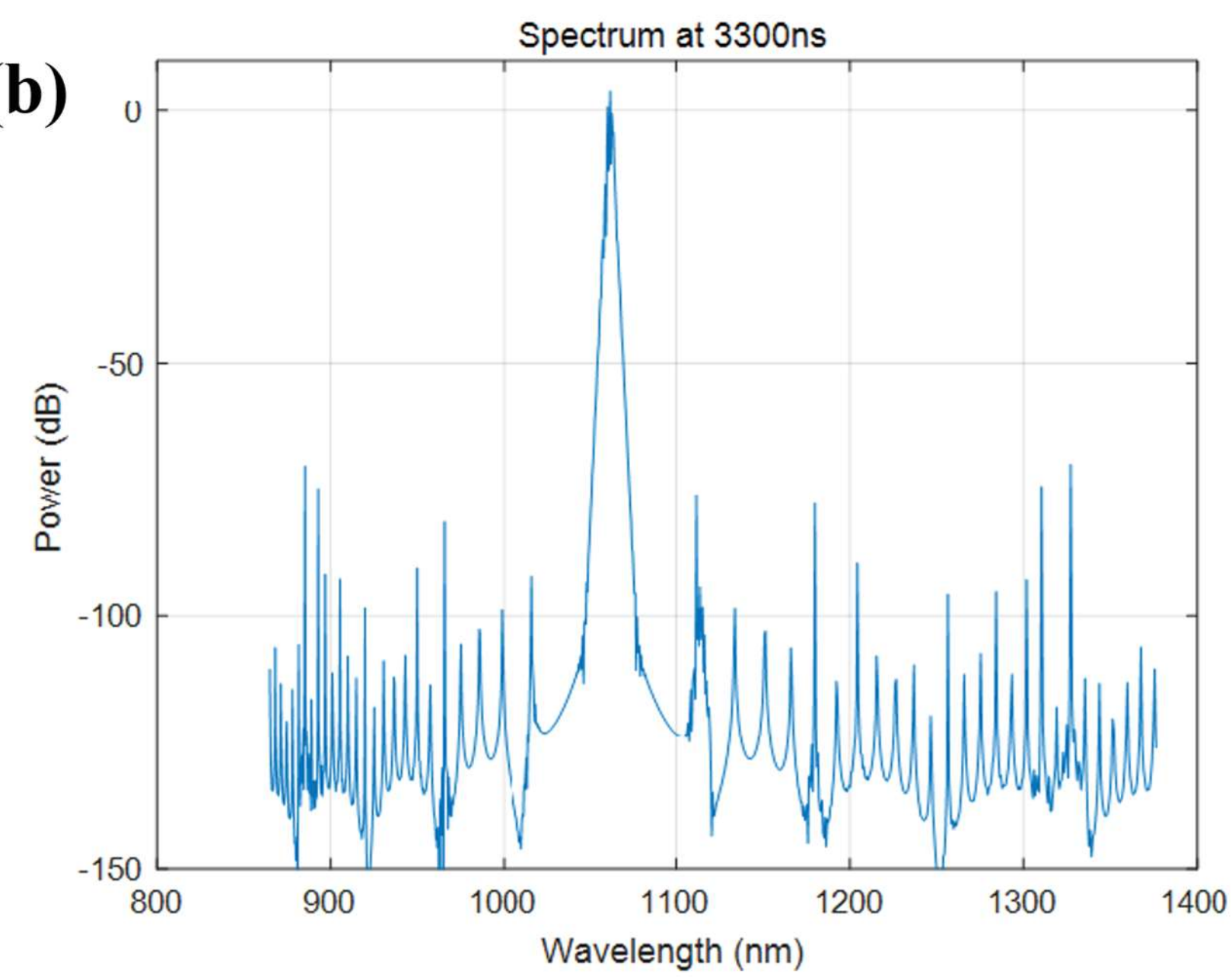

**Supplementary Figure 2. Simulation of microcomb generation using the Raman-modified Lugiato–Lefever equation.** (a) Simulated time-averaged intracavity power (blue) during a pump–resonance detuning scan, with the imposed detuning trajectory shown in red. (b) Simulated optical spectrum at 3300 ns, showing the pump and generated comb sidebands. The model includes Kerr nonlinearity, stimulated Raman scattering (SRS), cavity loss, pump-resonance detuning, external pumping, chromatic dispersion and gas-induced perturbation.

We modeled the intracavity comb dynamics using the Lugiato–Lefever equation (LLE). To describe comb generation in the silica microtoroid resonator, the LLE was modified to include Kerr nonlinearity, stimulated Raman scattering (SRS), cavity loss, pump-resonance detuning, external pumping, chromatic dispersion and gas-induced perturbation. The evolution of the intracavity electric field $E(t,\tau)$ is expressed as:

$$t_R \frac{\partial E}{\partial t} = -\left(\frac{\kappa}{2} + it_R\left(\omega_0 - \omega_p\right)\right)E + \sqrt{\frac{\kappa\eta}{\hbar\omega_0}}E_{in} - i\frac{1}{2}D_2\frac{\partial^2 E}{\partial \tau^2} + i\gamma L(1-f_R)|E|^2E + i\gamma L f_R(R * |E|^2)E + G_{gas}(E)E$$

Here, $t_R$ is the cavity round-trip time, $\kappa$ is the total cavity decay rate, $\omega_0$ and $\omega_p$ are the resonance and pump angular frequencies, respectively, and $E_{in}$ is the input pump field. The parameter $\eta$ describes the coupling coefficient, $D_2$ is the second-order dispersion coefficient, $\gamma$ is the nonlinear coefficient, $L$ is the cavity length, $f_R$ is the Raman fraction, and $R$ is the Raman response function. The convolution term $R * |E|^2$ accounts for the delayed Raman response. The additional term $G_{gas}(E)E$ represents gas-induced absorption and was used to simulate the attenuation of comb lines caused by molecular absorption.

Supplementary Fig. 2a shows the simulated time-averaged intracavity power during the pump-resonance detuning scan. The red line represents the detuning trajectory, while the blue curve shows the corresponding intracavity power. As the detuning is increased, the optical field becomes more efficiently coupled into the resonator and the time-averaged intracavity power increases. The oscillations and local fluctuations in the power trace arise from nonlinear intracavity dynamics during the detuning scan. This increased intracavity power provides the nonlinear condition required for Kerr and Raman-assisted comb generation.

Supplementary Fig. 2b shows the simulated optical spectrum at 3300 ns. A strong pump line is observed near the pump wavelength, accompanied by multiple sidebands extending across a broad spectral region. The asymmetric spectral extension toward longer wavelengths is consistent with stimulated Raman scattering in silica. This simulation confirms that the Raman-modified LLE can reproduce broadband microcomb formation before introducing gas-induced absorption.

We then used the same LLE framework to compare comb spectra before and after gas exposure. By introducing the gas-induced absorption term $G_{gas}(E)E$, the model predicts attenuation of selected comb-line intensities while maintaining the overall comb structure. These simulations provide the physical basis for using distributed comb-line intensity changes as sensing features in the main text.

## Baseline calibration and integrated comb-line intensity analysis

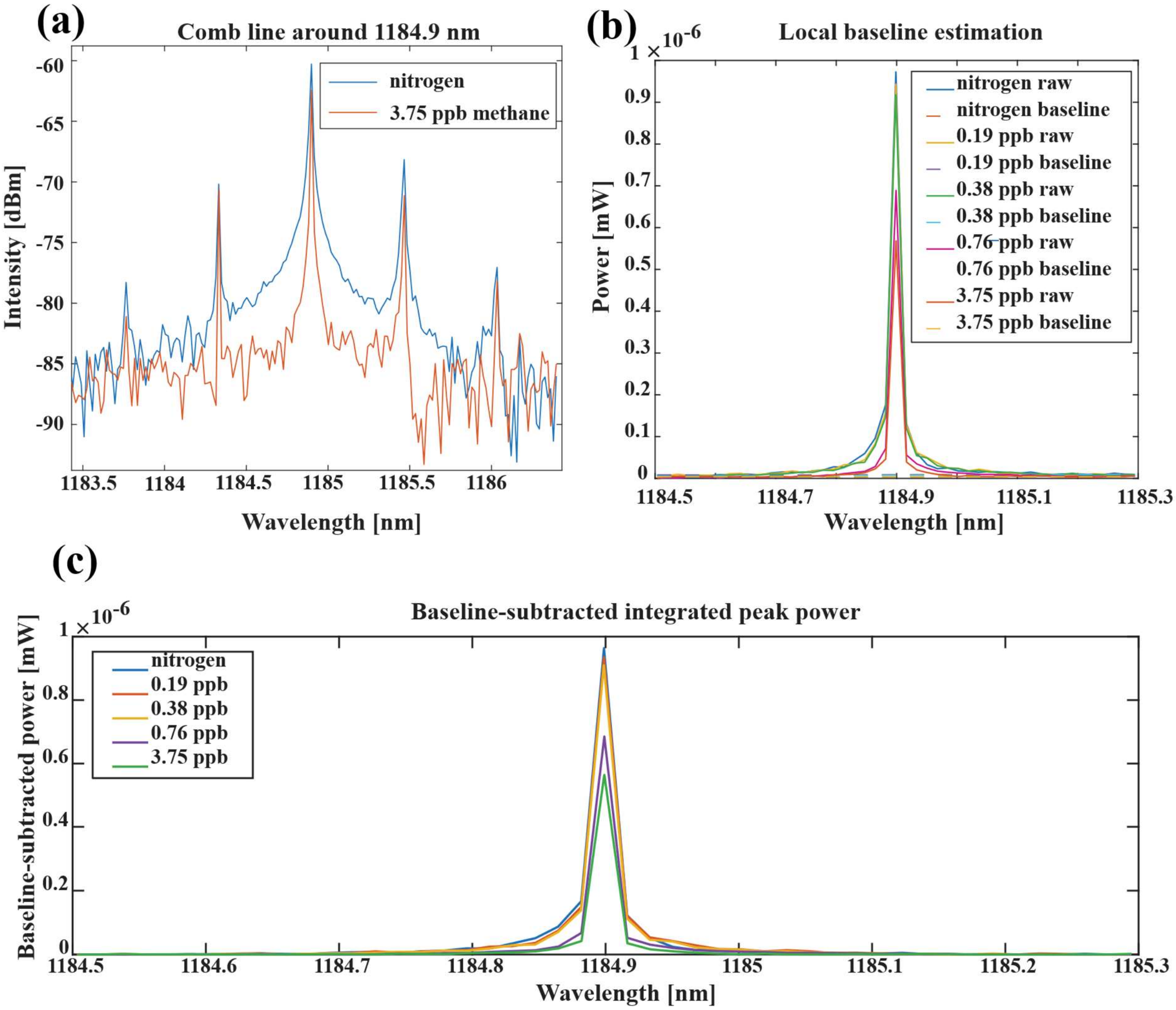

**Supplementary Figure 3. Local baseline correction and integrated-intensity analysis of a methane-sensitive comb line.** (a) Representative spectra near 1184.9 nm under nitrogen and 3.75 ppb methane, displayed in dBm. (b) Spectra converted to linear power units for nitrogen and methane concentrations of 0.19, 0.38, 0.76, and 3.75 ppb, together with local baselines estimated by first-order polynomial fits to the edges of the analysis window. (c) Corresponding baseline-subtracted spectra. The area under each corrected peak is evaluated by trapezoidal integration to quantify comb-line attenuation while accounting for changes in the local spectral background.

To quantify methane-induced attenuation of individual comb lines, we analyzed each optical spectrum analyzer (OSA) trace within a fixed wavelength window centered on the selected comb line. For a target comb wavelength $\lambda_0$, the local spectral region was defined as $\lambda_0 \pm 0.4$nm.

Because the OSA records spectral intensity in dBm, the intensity was first converted to linear power before baseline subtraction and numerical integration:

$$P_{\mathrm{mW}}(\lambda) = 10^{P_{\mathrm{dBm}}(\lambda)/10}.$$

This conversion is necessary because dBm is a logarithmic unit, and direct subtraction or integration in dBm can distort the extracted peak area.

For each selected wavelength window, the local baseline was estimated from the edge regions on both sides of the window. These edge-region data points were used to represent the local background or spectral pedestal around the comb line and were fitted with a first-order polynomial in linear power units:

$$P_{\mathrm{base}}(\lambda) = a\lambda + b.$$

The fitted baseline was then subtracted from the measured linear power spectrum:

$$P_{\mathrm{sub}}(\lambda) = P_{\mathrm{mW}}(\lambda) - P_{\mathrm{base}}(\lambda).$$

Negative values after baseline subtraction, mainly arising from noise fluctuations, were set to zero before integration. The baseline-subtracted peak height was defined as the maximum value of $P_{\mathrm{sub}}(\lambda)$. The integrated comb-line intensity was calculated by numerically integrating the baseline-subtracted power over the selected wavelength window:

$$A = \int_{\lambda_0-0.4}^{\lambda_0+0.4} P_{\mathrm{sub}}\,(\lambda)\, d\lambda.$$

In the implementation, this integral was evaluated using trapezoidal numerical integration. The resulting quantity has units of $\mathrm{mW}\cdot\mathrm{nm}$and represents the total local power associated with the selected comb-line feature after background removal.

For comparison between spectra before and after methane exposure, both the peak intensity change and the integrated-area change were calculated. The peak intensity change was obtained from the maximum OSA intensity in dBm:

$$\Delta P_{\mathrm{peak,dB}} = P_{\mathrm{peak,dBm}}^{\mathrm{CH_4}} - P_{\mathrm{peak,dBm}}^{\mathrm{N_2}}.$$

The integrated-area change was calculated in dB from the ratio of the baseline-subtracted integrated areas:

$$\Delta A_{\mathrm{dB}} = 10\log_{10}\left(\frac{A_{\mathrm{CH_4}}}{A_{\mathrm{N_2}}}\right).$$

Here, $A_{N_2}$ and $A_{CH_4}$ are the integrated comb-line areas measured before and after methane introduction, respectively. Compared with single-point peak-height analysis, the integrated-area metric is less sensitive to local sampling fluctuations in the OSA trace and provides a more robust estimate of the total attenuation of the comb-line feature.

This baseline-calibrated analysis was also used to distinguish true intensity attenuation from changes in the local spectral pedestal. Because the apparent width and shape of an OSA-resolved comb line can be affected by the local baseline, noise floor, and finite OSA resolution, the baseline-subtracted integrated power was used as a complementary metric to the peak intensity for evaluating methane-induced spectral changes.

## Comparison between PCA and PLS-DA

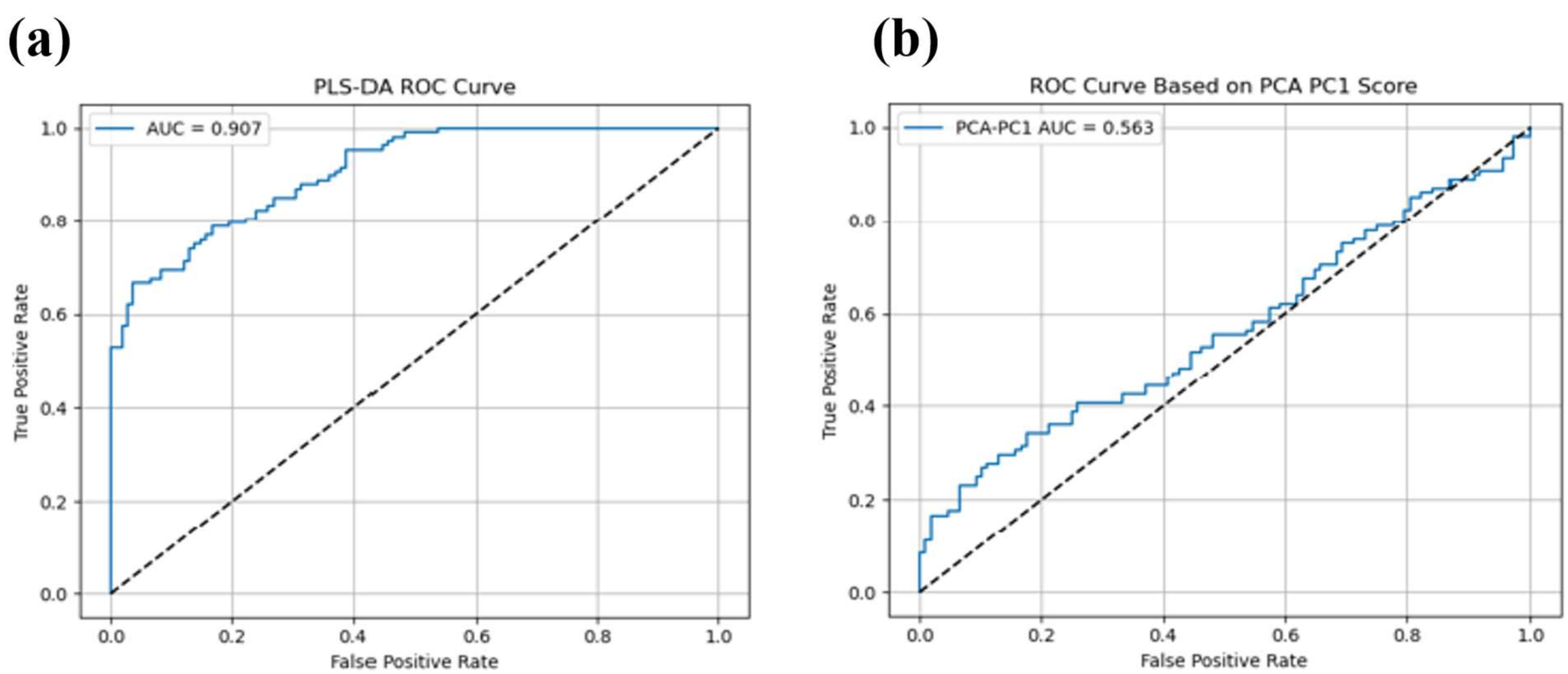

**Supplementary Figure 4. Comparison of supervised and unsupervised spectral discrimination of methane exposure.** (a) Receiver operating characteristic (ROC) curves obtained using partial least-squares discriminant analysis (PLS-DA). (b) the first principal-component score from principal component analysis (PCA). Both analyses use the same dataset comprising 10 min of nitrogen exposure and 10 min of 0.19 ppb methane exposure. The areas under the ROC curves are 0.907 for PLS-DA and 0.563 for PCA. Dashed diagonal lines indicate chance-level discrimination.

We compared principal component analysis (PCA) and partial least-squares discriminant analysis (PLS-DA) to evaluate their ability to distinguish methane exposure from the nitrogen baseline. PCA is an unsupervised dimensionality-reduction method that projects high-dimensional spectral data into a lower-dimensional space by maximizing the variance in the dataset, without using class-label information. In contrast, PLS-DA is a supervised classification method that incorporates class labels during model training and identifies latent variables that maximize the separation between predefined groups.

For the same dataset, consisting of 10 min of nitrogen exposure and 10 min of 0.19 ppb methane exposure, PLS-DA exhibited a higher area under the ROC curve (AUC) than PCA. This result indicates that the supervised PLS-DA model provides stronger discrimination between nitrogen and methane conditions than PCA for this dataset. The improved performance is expected because PLS-DA uses class information to extract methane-relevant spectral features, whereas PCA primarily captures the directions of largest overall variance, which may include noise, baseline drift, or other variations unrelated to methane detection.

Therefore, PLS-DA was selected as the primary analysis method for methane classification in this work.

## Negative control

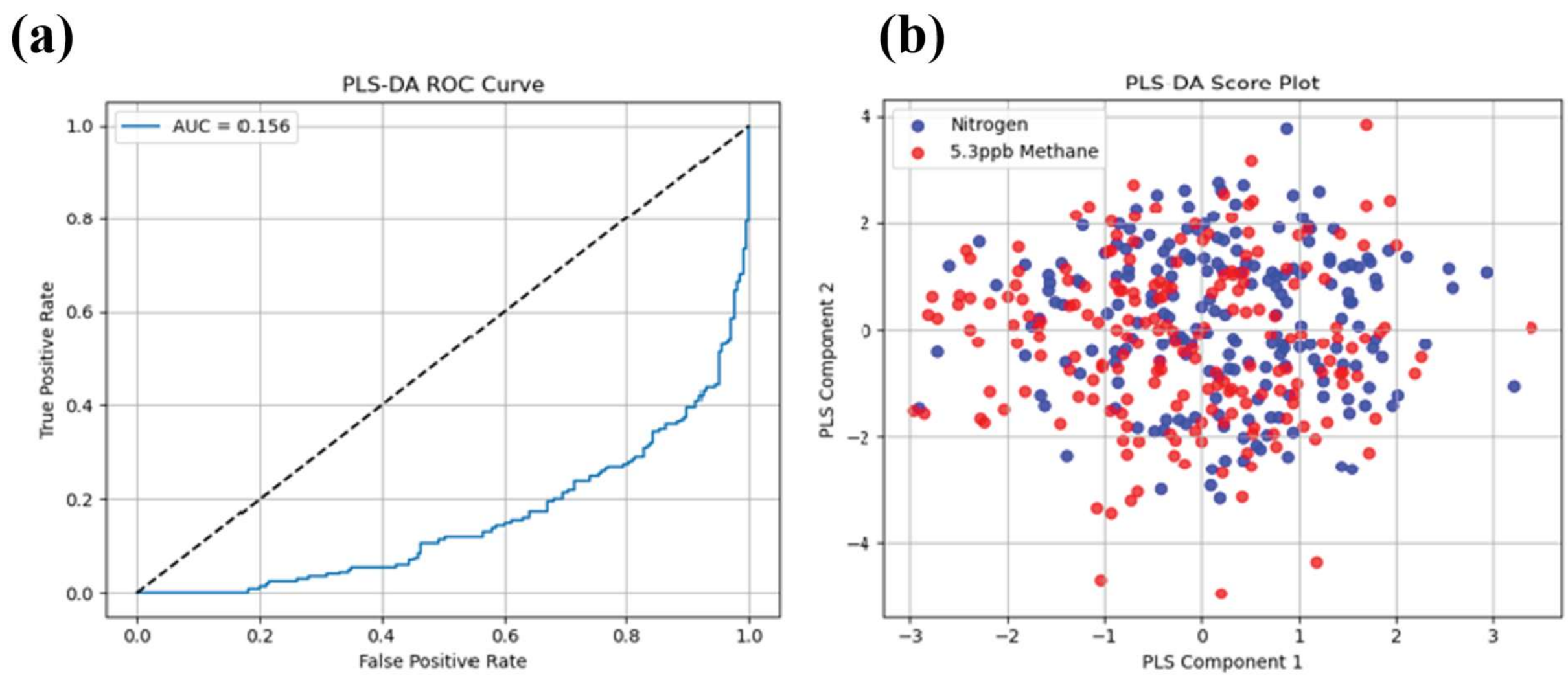

**Supplementary Figure 5. Nitrogen-only control for PLS-DA classification.** Spectra acquired during 20 min of nitrogen exposure were divided into two groups and assigned nominal nitrogen and methane labels, although no methane was introduced. (a) Receiver operating characteristic curve obtained using the assigned labels, with an area under the curve of 0.156; the dashed diagonal indicates chance-level discrimination. (b) PLS-DA score plot showing overlap between the two nitrogen-only groups. The reported classification accuracy was approximately 50%.

A negative control experiment was performed to evaluate whether the PLS-DA model could generate false-positive methane detection results from nitrogen-only data. In this experiment, the system was exposed to nitrogen for 20 min. The acquired nitrogen data were then divided into two groups: one group was labeled as the nitrogen baseline, while the other group was intentionally mislabeled as methane. Because both groups were collected under the same nitrogen environment, no true methane-induced spectral difference should exist between them.

The PLS-DA analysis produced a classification accuracy of approximately 50%, which corresponds to random classification between the two artificially assigned groups. The area under the ROC curve (AUC) was below 0.156, further confirming that the model did not identify a meaningful separation between the two nitrogen-only datasets. These results indicate that the PLS-DA model does not mistakenly classify nitrogen baseline fluctuations as methane response and does not produce false-positive methane detection under this negative control condition.

## Variable importance in projection analysis

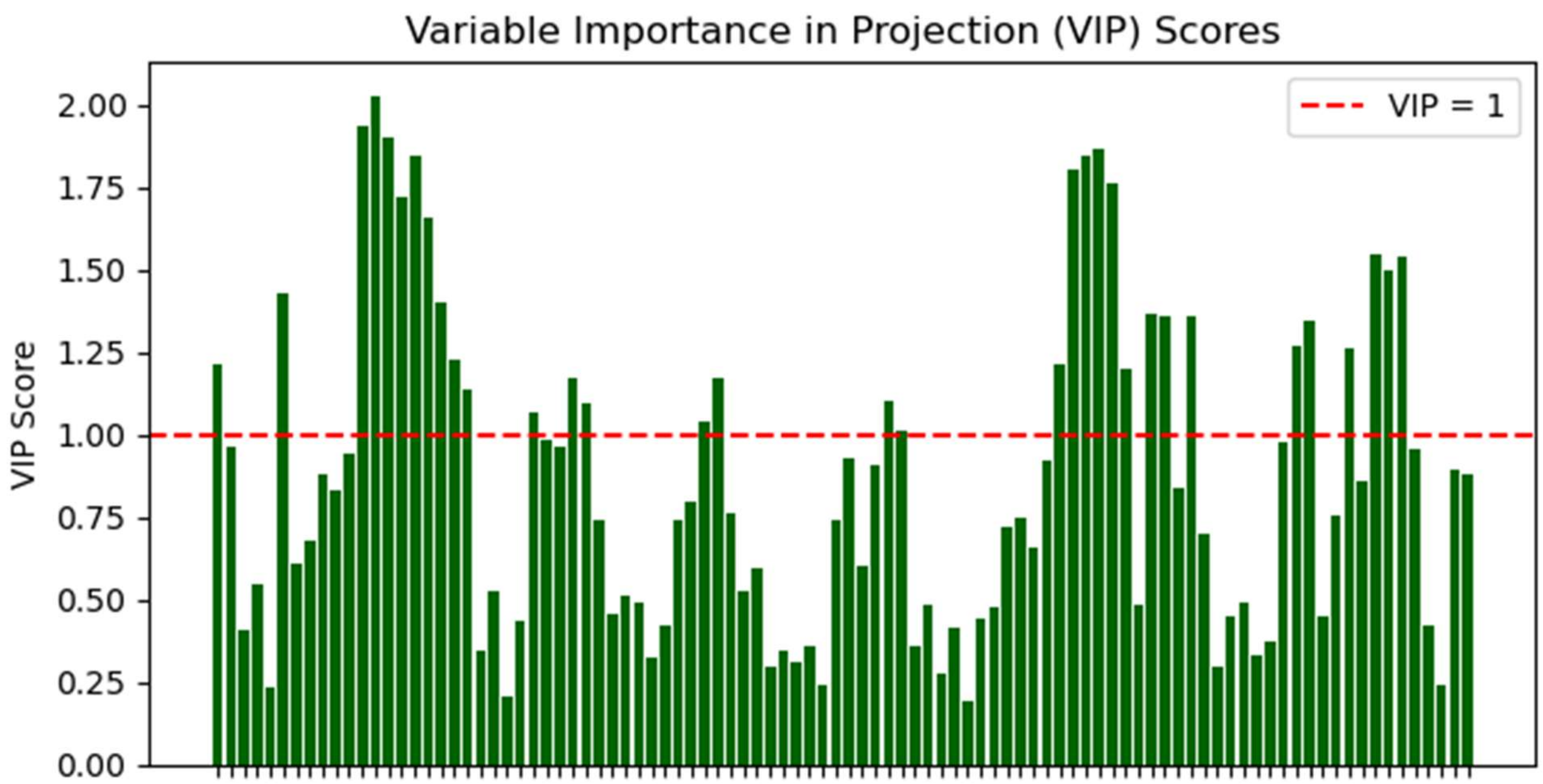

**Supplementary Figure 6. Contributions of individual comb lines to methane classification.** Variable importance in projection (VIP) scores for the comb-line features used in the partial least-squares discriminant analysis model. Each bar represents one comb line, indexed along the horizontal axis. The red dashed line marks VIP = 1, a commonly used reference for identifying influential variables. Multiple comb lines exceed this reference, indicating that the model draws on spectral information distributed across several channels.

Variable importance in projection (VIP) scores were used to evaluate the relative contribution of each comb line to the PLS-DA classification model. In PLS-DA, VIP scores summarize how strongly each input variable contributes to the latent variables that separate different classes. Therefore, in our analysis, the VIP score provides a quantitative measure of which comb lines are most important for distinguishing methane exposure from the nitrogen baseline.

The VIP-score analysis supports the multichannel nature of the microcomb sensing response. Although the contribution of each comb line was not identical, the classification was not determined by a single dominant wavelength. Instead, multiple comb lines contributed to the separation between nitrogen and methane spectra, consistent with the improved classification performance observed when additional comb lines were included in the analysis. Comb lines located near the center of the methane absorption region generally exhibited higher VIP scores. In addition, comb lines with stronger intensity and higher temporal stability also tended to contribute more significantly to the model. This trend is physically reasonable because comb lines overlapping with methane absorption features are more sensitive to methane-induced attenuation, while stable and high-intensity comb lines provide more reliable spectral features for classification.

Therefore, the VIP-score analysis provides two complementary insights. First, it confirms that methane detection arises from distributed spectral information encoded across multiple comb lines, rather than from an isolated spectral channel. Second, it provides useful guidance for feature selection by identifying the most informative and reliable comb lines. Prioritizing high-VIP-score channels may further improve classification accuracy, reduce the influence of noisy or less relevant comb lines, and simplify the sensing model while preserving its methane-detection capability.

## Low-concentration validation of the PLS-DA-based methane detection threshold.

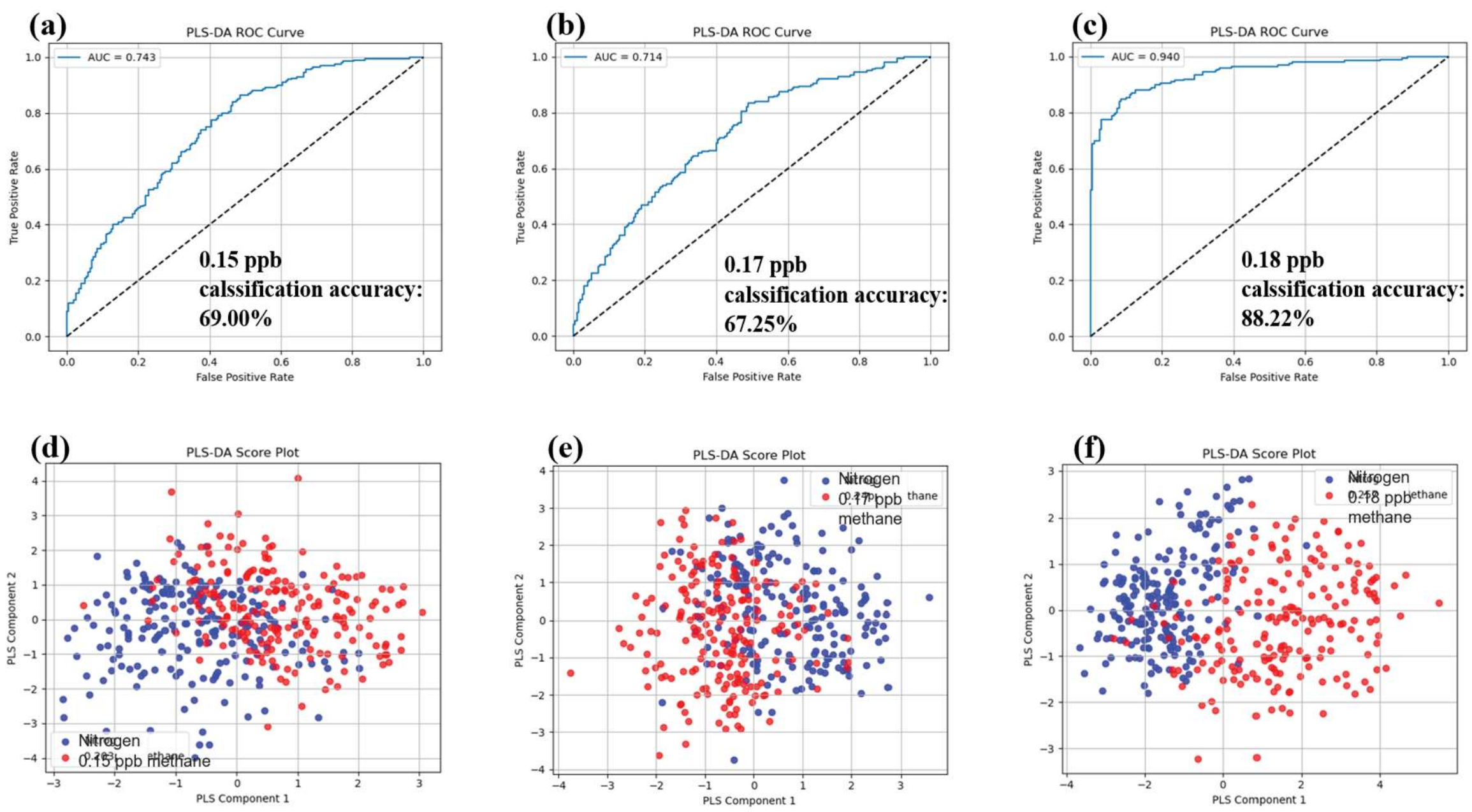

**Supplementary Figure 7. Experimental evaluation of methane discrimination near the detection threshold.** Receiver operating characteristic curves (a, b, c) and corresponding partial least-squares discriminant analysis score plots (d, e, f) for methane concentrations of 0.15, 0.17, and 0.18 ppb, from left to right, compared with nitrogen baseline spectra. The respective areas under the ROC curves are 0.743, 0.714, and 0.940, with classification accuracies of 69.00%, 67.25%, and 88.22%. Only 0.18 ppb methane exceeds the predefined 80% classification-accuracy criterion, supporting an experimentally demonstrated detection threshold of approximately 180 ppt. Dashed diagonal lines in the ROC plots indicate chance-level discrimination.

To validate the methane limit of detection obtained from the concentration-dependent classification and response analyses, we performed additional experiments at methane concentrations close to the estimated detection threshold. Methane concentrations of 0.15 ppb, 0.17 ppb, and 0.18 ppb were tested and compared against nitrogen baseline spectra using partial least-squares discriminant analysis (PLS-DA). For each concentration, receiver operating characteristic (ROC) analysis and PLS-DA score plots were used to evaluate the separability between nitrogen and methane-exposed spectra.

For 0.15 ppb methane, the PLS-DA model produced an AUC of 0.743 and a classification accuracy of 69.00%. For 0.17 ppb methane, the AUC was 0.714 and the classification accuracy was 67.25%. These accuracies were below the predefined 80% accuracy threshold, indicating that methane could not be reliably distinguished from nitrogen at these concentrations under the present experimental conditions. By contrast, for 0.18 ppb methane, the AUC increased to 0.940 and the classification accuracy reached 88.22%, exceeding the 80% detection criterion. The improved ROC performance and clearer separation in the PLS-DA score plot indicate that methane-induced spectral perturbations become reliably distinguishable at this concentration.

Based on the 80% classification-accuracy criterion, the direct low-concentration experiment gives an experimentally verified LOD of approximately 180 ppt. This experimentally determined value is consistent with the LODs estimated from the classification-accuracy fitting and the signal-response calibration in Fig. 4(e), (f). The agreement among these three analyses confirms that the reported sub-ppb detection limit is robust and supported by both model-based fitting and direct experimental validation.

## Reproducibility of microcomb generation

Raman-assisted microcomb generation is highly reproducible, with three characteristic spectral regions consistently appearing across independent measurements. These regions correspond to the first-, second-, and third-order Raman combs, confirming the repeatable formation of cascaded Raman components. The consistent spectral coverage within the methane absorption region (1100–1200 nm) provides a reliable foundation for broadband molecular sensing. Furthermore, the reproducibility of these spectral features opens opportunities for developing more advanced machine-learning models that exploit the collective response of multiple comb lines. Beyond binary classification of gas presence or absence, future models could enable quantitative prediction of gas concentrations, identification of multiple gas species, and more sophisticated analysis of complex gas mixtures, extending the capabilities of microcomb-based sensing platforms.

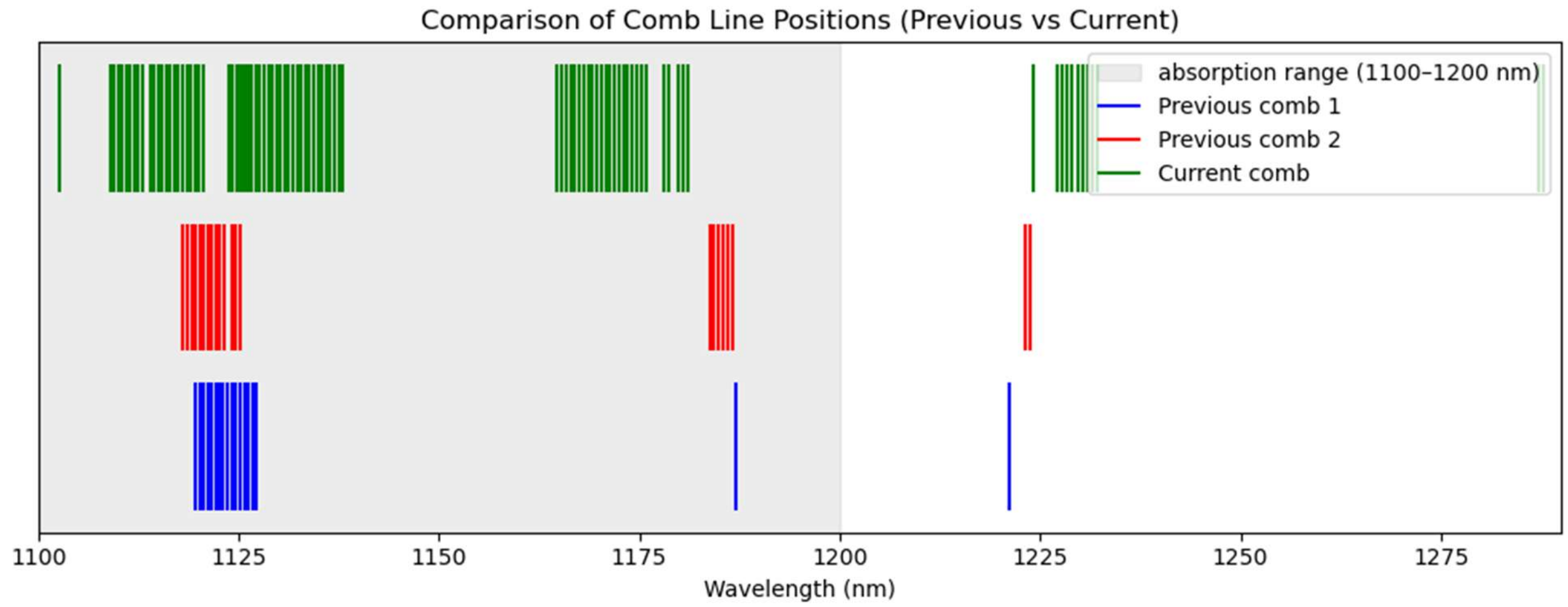

**Supplementary Figure 8. Comparison of comb-line positions across independent microcomb measurements.** Wavelength positions of comb lines from three independently acquired spectra, shown as blue, red, and green vertical markers. The shaded region denotes the 1100–1200 nm methane absorption window considered in this study. The comparison shows recurring groups of comb lines and repeated spectral coverage within the sensing window, together with variations in individual line positions and the number of generated lines.

## References

1. US EPA, O. Understanding Global Warming Potentials. https://www.epa.gov/ghgemissions/understanding-global-warming-potentials (2016).
2. Canadell, J. G. *et al.* Global carbon and other biogeochemical cycles and feedbacks. in *Climate Change 2021: The Physical Science Basis. Contribution of Working Group I to the Sixth Assessment Report of the Intergovernmental Panel on Climate Change* (eds Masson-Delmotte, V. et al.) 673–816 (Cambridge University Press, Cambridge, United Kingdom and New York, NY, USA, 2021). doi:10.1017/9781009157896.001.
3. US EPA, O. Importance of Methane. https://www.epa.gov/gmi/importance-methane (2016).
4. Erland, B. M., Thorpe, A. K. & Gamon, J. A. Recent Advances Toward Transparent Methane Emissions Monitoring: A Review. *Environ. Sci. Technol.* **56**, 16567–16581 (2022).
5. Aldhafeeri, T., Tran, M.-K., Vrolyk, R., Pope, M. & Fowler, M. A Review of Methane Gas Detection Sensors: Recent Developments and Future Perspectives. *Inventions* **5**, 28 (2020).
6. Kassi, S., Gao, B., Romanini, D. & Campargue, A. The near-infrared (1.30–1.70 µm) absorption spectrum of methane down to 77 K. *Phys. Chem. Chem. Phys.* **10**, 4410–4419 (2008).
7. Smith, B. C. The Big Review II: The Physical Mechanism of Infrared Absorbance and Peak Types | Spectroscopy Online. https://www.spectroscopyonline.com/view/the-big-review-ii-the-physical-mechanism-of-infrared-absorbance-and-peak-types (2026).
8. Lin, S., Chang, J., Sun, J. & Xu, P. Improvement of the Detection Sensitivity for Tunable Diode Laser Absorption Spectroscopy: A Review. *Front. Phys.* **10**, (2022).
9. O'Keefe, A. & Deacon, D. A. G. Cavity ring-down optical spectrometer for absorption measurements using pulsed laser sources. *Rev. Sci. Instrum.* **59**, 2544–2551 (1988).

10. Crosson, E. R. A cavity ring-down analyzer for measuring atmospheric levels of methane, carbon dioxide, and water vapor. *Appl. Phys. B* **92**, 403–408 (2008).
11. Engeln, R., Berden, G., Peeters, R. & Meijer, G. Cavity enhanced absorption and cavity enhanced magnetic rotation spectroscopy. *Rev. Sci. Instrum.* **69**, 3763–3769 (1998).
12. Barry, H. R., Corner, L., Hancock, G., Peverall, R. & Ritchie, G. A. D. Cavity-enhanced absorption spectroscopy of methane at 1.73 µm. *Chem. Phys. Lett.* **333**, 285–289 (2001).
13. Xiao, H. *et al.* Ultra-sensitive ppb-level methane detection based on NIR all-optical photoacoustic spectroscopy by using differential fiber-optic microphones with gold-chromium composite nanomembrane. *Photoacoustics* **26**, 100353 (2022).
14. Bauer, C. P. *et al.* High-sensitivity dual-comb and cross-comb spectroscopy across the infrared using a widely tunable and free-running optical parametric oscillator. *Nat. Commun.* **15**, 7211 (2024).
15. Baumann, E. *et al.* Spectroscopy of the methane ${\ensuremath{\nu}}_{3}$ band with an accurate midinfrared coherent dual-comb spectrometer. *Phys. Rev. A* **84**, 062513 (2011).
16. Ma, Y., Liu, Y., He, Y., Qiao, S. & Sun, H. Design of multipass cell with dense spot patterns and its performance in a light-induced thermoelastic spectroscopy-based methane sensor. *Light Adv. Manuf.* **6**, 5–13 (2025).
17. Gu, C. *et al.* Passive coherent dual-comb spectroscopy based on optical-optical modulation with free running lasers. *PhotoniX* **1**, 7 (2020).
18. Adler, F., Thorpe, M. J., Cossel, K. C. & Ye, J. Cavity-Enhanced Direct Frequency Comb Spectroscopy: Technology and Applications. *Annu. Rev. Anal. Chem.* **3**, 175–205 (2010).
19. Coddington, I., Newbury, N. & Swann, W. Dual-comb spectroscopy. *Optica* **3**, 414–426 (2016).
20. Ghosh, R., Aslam, M. & Kalita, H. Graphene derivatives for chemiresistive gas sensors: A review. *Mater. Today Commun.* **30**, 103182 (2022).
21. Recum, P. & Hirsch, T. Graphene-based chemiresistive gas sensors. *Nanoscale Adv.* **6**, 11–31 (2024).
22. Tang, R., Shi, Y., Hou, Z. & Wei, L. Carbon Nanotube-Based Chemiresistive Sensors. *Sensors* **17**, 882 (2017).
23. Zhang, Y. *et al.* Recent Progress in Low-Power-Consumption Metal Oxide Semiconductor Gas Sensors. *Materials* **18**, 4864 (2025).
24. Xiao, L., Xu, S., Yu, G. & Liu, S. Efficient hierarchical mixed Pd/SnO2 porous architecture deposited microheater for low power ethanol gas sensor. *Sens. Actuators B Chem.* **255**, 2002–2010 (2018).
25. Wang, Y. & Zhou, Y. Recent Progress on Anti-Humidity Strategies of Chemiresistive Gas Sensors. *Materials* **15**, 8728 (2022).
26. Kim, K. *et al.* Synergistic approach to simultaneously improve response and humidity-independence of metal-oxide gas sensors. *J. Hazard. Mater.* **424**, 127524 (2022).
27. Mahdavi, H., Rahbarpour, S., Hosseini-Golgoo, S.-M. & Jamaati, H. Reducing the destructive effect of ambient humidity variations on gas detection capability of a temperature modulated gas sensor by calcium chloride. *Sens. Actuators B Chem.* **331**, 129091 (2021).
28. Gao, Z. *et al.* A facile PDMS coating approach to room-temperature gas sensors with high humidity resistance and long-term stability. *Sens. Actuators B Chem.* **325**, 128810 (2020).
29. Su, J. Label-Free Biological and Chemical Sensing Using Whispering Gallery Mode Optical Resonators: Past, Present, and Future. *Sensors* **17**, 540 (2017).

30. Foreman, M. R., Swaim, J. D. & Vollmer, F. Whispering gallery mode sensors. *Adv. Opt. Photonics* **7**, 168–240 (2015).
31. Jiang, X., Qavi, A. J., Huang, S. H. & Yang, L. Whispering-Gallery Sensors. *Matter* **3**, 371–392 (2020).
32. Li, C. *et al.* Part-per-Trillion Trace Selective Gas Detection Using Frequency Locked Whispering-Gallery Mode Microtoroids. *ACS Appl. Mater. Interfaces* **14**, 42430–42440 (2022).
33. Suebka, S., McLeod, E. & Su, J. Ultra-high-Q free-space coupling to microtoroid resonators. *Light Sci. Appl.* **13**, 75 (2024).
34. Xu, Y. *et al.* Low Part-Per-Trillion, Humidity Resistant Detection of Nitric Oxide Using Microtoroid Optical Resonators. *ACS Appl. Mater. Interfaces* **16**, 5120–5128 (2024).
35. Choi, G., Gin, A. & Su, J. Optical frequency combs in aqueous and air environments at visible to near-IR wavelengths. *Opt. Express* **30**, 8690–8699 (2022).
36. Kippenberg, T. J., Holzwarth, R. & Diddams, S. A. Microresonator-Based Optical Frequency Combs. *Science* **332**, 555–559 (2011).
37. Del'Haye, P. *et al.* Optical frequency comb generation from a monolithic microresonator. *Nature* **450**, 1214–1217 (2007).
38. Okawachi, Y. *et al.* Competition between Raman and Kerr effects in microresonator comb generation. *Opt. Lett.* **42**, 2786–2789 (2017).
39. Fujii, S., Kato, T., Suzuki, R., Hori, A. & Tanabe, T. Transition between Kerr comb and stimulated Raman comb in a silica whispering gallery mode microcavity. *JOSA B* **35**, 100–106 (2018).
40. Cherenkov, A. V. *et al.* Raman-Kerr frequency combs in microresonators with normal dispersion. *Opt. Express* **25**, 31148–31158 (2017).
41. Picqué, N. & Hänsch, T. W. Frequency comb spectroscopy. *Nat. Photonics* **13**, 146–157 (2019).
42. Weichman, M. L. *et al.* Broadband molecular spectroscopy with optical frequency combs. *J. Mol. Spectrosc.* **355**, 66–78 (2019).
43. Choi, G., Gin, A., Su, J. & Su, J. Optical frequency combs in aqueous and air environments at visible to near-IR wavelengths. *Opt. Express* **30**, 8690–8699 (2022).
44. Choi, G. & Su, J. Impact of stimulated Raman scattering on dark soliton generation in a silica microresonator. *J. Phys. Photonics* **5**, 014001 (2022).
45. Suebka, S., Gin, A. & Su, J. Frequency locked whispering evanescent resonator (FLOWER) for biochemical sensing applications. *Nat. Protoc.* **20**, 1616–1650 (2025).
46. Gin, A. *et al.* Label-free, real-time monitoring of membrane binding events at zeptomolar concentrations using frequency-locked optical microresonators. *Nat. Commun.* **15**, 7445 (2024).
47. Gin, A., Nguyen, P.-D., Serrano, G., Alexander, G. E. & Su, J. Towards early diagnosis and screening of Alzheimer's disease using frequency locked whispering gallery mode microtoroids. *Npj Biosensing* **1**, 1–9 (2024).
48. Li, C. *et al.* Part-per-Trillion Trace Selective Gas Detection Using Frequency Locked Whispering-Gallery Mode Microtoroids. *ACS Appl. Mater. Interfaces* **14**, 42430–42440 (2022).
49. Xu, Y. *et al.* Free-Space-Coupled Frequency-Locked Microtoroid Resonators With Reactive Polymer Functionalization for Part-Per-Trillion Gas Detection. *Laser Photonics Rev.* **20**, e71128 (2026).

50. Kim, S.-K. *et al.* Methotrexate Inhibits the Binding of the Severe Acute Respiratory Syndrome Coronavirus 2 (SARS-CoV-2) Receptor Binding Domain to the Host-Cell Angiotensin-Converting Enzyme-2 (ACE-2) Receptor. *ACS Pharmacol. Transl. Sci.* **7**, 348–362 (2024).
51. Young Yang, M. *et al.* Agonist activation to open the Gα subunit of the GPCR–G protein precoupled complex defines functional agonist activation of TAS2R5. *Proc. Natl. Acad. Sci.* **121**, e2409987121 (2024).
52. Su, J., Liggett, S. B., Kim, S.-K., Kim, D. & Goddard, W. A. Accelerating GPCR drug discovery through computation and experiment integrated with direct detection of ligand binding events. *Npj Drug Discov.* **3**, 13 (2026).
53. Zhang, Y. & Su, J. FLOWER: a frequency-locked optical whispering evanescent resonator for label-free molecular detection. *Chem. Commun.* **61**, 13561–13570 (2025).
54. Chen, L., Li, C., Liu, Y., Su, J. & McLeod, E. Three-Dimensional Simulation of Particle-Induced Mode Splitting in Large Toroidal Microresonators. *Sensors* **20**, 5420 (2020).
55. Chembo, Y. K., Grudinin, I. S. & Yu, N. Spatiotemporal dynamics of Kerr-Raman optical frequency combs. *Phys. Rev. A* **92**, 043818 (2015).
56. Chembo, Y. K. & Menyuk, C. R. Spatiotemporal Lugiato-Lefever formalism for Kerr-comb generation in whispering-gallery-mode resonators. *Phys. Rev. A* **87**, 053852 (2013).
57. Wong, F. *et al.* An explainable deep learning platform for molecular discovery. *Nat. Protoc.* **20**, 1020–1056 (2025).
58. Liang, Q. *et al.* Breath analysis by ultra-sensitive broadband laser spectroscopy detects SARS-CoV-2 infection. *J. Breath Res.* **17**, 036001 (2023).
59. Müller, M. *et al.* Comparison of photoacoustic spectroscopy and cavity ring-down spectroscopy for ambient methane monitoring at Hohenpeißenberg. *Atmos. Meas. Tech.* **16**, 4263–4270 (2023).
60. Mu, T. *et al.* Sub-ppb Methane Detection via EMD–Wavelet Adaptive Thresholding in Wavelength Modulation TDLAS: A Hybrid Denoising Approach for Trace Gas Sensing. *Sensors* **25**, 5167 (2025).
61. Chen, Y. *et al.* Rapid ppb-Level Methane Detection Based on Quartz-Enhanced Photoacoustic Spectroscopy. *Anal. Chem.* **97**, 6780–6787 (2025).
62. Mikkonen, T., Hieta, T., Genty, G. & Toivonen, J. Sensitive multi-species photoacoustic gas detection based on mid-infrared supercontinuum source and miniature multipass cell. *Phys. Chem. Chem. Phys.* **24**, 19481–19487 (2022).
63. Rieker, G. B. *et al.* Frequency Comb-Based Remote Sensing of Greenhouse Gases over Kilometer Air Paths. Preprint at https://doi.org/10.48550/arXiv.1406.3326 (2014).
64. Zhu, Z., Xu, Y. & Jiang, B. A One ppm NDIR Methane Gas Sensor with Single Frequency Filter Denoising Algorithm. *Sensors* **12**, 12729–12740 (2012).
65. Kimiagar, S., Najafi, V., Witkowski, B., Pietruszka, R. & Godlewski, M. High performance and low temperature coal mine gas sensor activated by UV-irradiation. *Sci. Rep.* **8**, 16298 (2018).
66. Ultrasensitive Electrochemical Methane Sensors Based on Solid Polymer Electrolyte-Infused Laser-Induced Graphene | ACS Applied Materials & Interfaces | ACS Publications. https://pubs.acs.org/aamick/article-abstract/11/6/6166/1277512/Ultrasensitive-Electrochemical-Methane-Sensors?redirectedFrom=fulltext.
67. Pellistor (Catalytic bead sensor) for methane and combustible gas. *Nano Environmental Technology S.r.l.* https://www.nenvitech.com/products/catalytic-bead-sensors/.